\documentclass{article}

\usepackage{arxiv}      

\usepackage[utf8]{inputenc} 
\usepackage[T1]{fontenc}    

\usepackage{amsmath, amssymb, amsfonts, amsthm} 
\usepackage{mathtools}     
\usepackage{bm}            
\usepackage{mathrsfs}      

\usepackage{graphicx}      
\usepackage{subfig}        
\usepackage{float}         

\usepackage{booktabs}      
\usepackage{multirow}      
\usepackage{makecell}      

\usepackage{algorithm}
\usepackage{algorithmicx}
\usepackage{algpseudocode}

\usepackage[numbers]{natbib} 
\usepackage{doi}            

\usepackage{hyperref}
\usepackage{url}

\usepackage{xcolor}         
\usepackage{textcomp}       
\usepackage{manyfoot}       
\usepackage{lipsum}         
\usepackage{microtype}      
\usepackage{nicefrac}       

\usepackage[title]{appendix}

\usepackage{lineno}         

\title{\textit{ShapeGen3DCP}: A Deep Learning Framework for Layer Shape Prediction in 3D Concrete Printing}

\author{Giacomo Rizzieri$^{1,*}$, Federico Lanteri$^{1}$, Liberato Ferrara$^{1}$, Massimiliano Cremonesi$^{1}$\\
	$^{1}$Department of Civil and Environmental Engineering, Politecnico di Milano\\
    Piazza Leonardo da Vinci, 32, 20133, Milano, Italy\\
	$^{*}$\texttt{giacomo.rizzieri@polimi.it} \\
}

\begin{document}
\maketitle

\begin{abstract}
This work introduces \textit{ShapeGen3DCP}, a deep learning framework for fast and accurate prediction of filament cross-sectional geometry in 3D Concrete Printing (3DCP). The method is based on a neural network architecture that takes as input both material properties in the fluid state (density, yield stress, plastic viscosity) and process parameters (nozzle diameter, nozzle height, printing and flow velocities) to directly predict extruded layer shapes. To enhance generalization, some inputs are reformulated into dimensionless parameters that capture underlying physical principles. Predicted geometries are compactly represented using Fourier descriptors, which enforce smooth, closed, and symmetric profiles while reducing the prediction task to a small set of coefficients. The training dataset was synthetically generated using a well-established Particle Finite Element (PFEM) model of 3DCP, overcoming the scarcity of experimental data. Validation against diverse numerical and experimental cases shows strong agreement, confirming the framework’s accuracy and reliability. This opens the way to practical uses ranging from pre-calibration of print settings, minimizing or even eliminating trial-and-error adjustments, to toolpath optimization for more advanced designs. Looking ahead, coupling the framework with simulations and sensor feedback could enable closed-loop digital twins for 3DCP, driving real-time process optimization, defect detection, and adaptive control of printing parameters.
\end{abstract}

\keywords{Additive manufacturing \and 3D concrete printing (3DCP) \and Filament geometry \and Machine learning \and Artificial Neural Networks (ANNs) \and Fourier descriptors}

\section{Introduction}\label{sec:intro}
Additive manufacturing with cement-based materials, particularly 3D Concrete Printing (3DCP), has attracted increasing attention over the past decade as a means to enhance productivity, performance, and aesthetics in the construction industry. By extruding a cementitious mortar through a digitally controlled nozzle, 3DCP enables the layer-by-layer fabrication of structures without the need for formwork, offering greater design flexibility while reducing material waste and construction time.

Since in 3DCP quality control begins at the scale of the single filament \cite{wolfs2021}, achieving high-quality and consistent shapes depends above all on precise material control. The cementitious "ink" must meet stringent rheological and mechanical requirements \cite{buswell2018}, guaranteeing sufficient extrudability to flow smoothly through the nozzle and adequate buildability to retain its shape and support subsequent layers.

Despite significant progress in characterizing the fresh properties of 3D-printable cementitious materials also through innovative rheological tests (e.g., flow table test \cite{tay2019}, slug test \cite{ducoulombier2021}), controlling and predicting the printing process remains a major challenge. This is because the outcome depends not only on material parameters, but also on numerous printing variables, including nozzle geometry, nozzle-to-substrate distance, extrusion rate, printing speed, and toolpath strategy. 

As recently demonstrated by insights from full-scale 3DCP pilot projects \cite{bos2022}, process parameters must be carefully tuned in accordance with the material’s rheological properties, ambient conditions, and specific design requirements (e.g., surface finish quality and interlayer adhesion). Moreover, the strong interdependence among the aforesaid factors adds significant complexity to 3DCP, posing challenges to the establishment of standardized procedures and regulatory frameworks. As a result, practitioners often rely on empirical expertise and time-consuming trial-and-error approaches to achieve consistent print quality.

\subsection{Numerical methods in 3DCP}
Numerical methods are being developed to support designers and offer predictive insight into the 3D printing process. The most established approach relies on solid FEM combined with selective layer or element activation to predict failure modes such as buckling or plastic collapse and estimate buildabuility \cite{wolfs2018, ooms2021}. Though recent studies have demonstrated the applicability of this technique also to complex geometries \cite{nguyen-van2021, rymes2023} or in combination with advanced constitutive laws \cite{chen2025}, most models still idealize the filament geometry as rectangular, overlooking the true cross-sectional shape. However, it was demonstrated in \cite{liu2021, reinold2024} that this simplification can substantially impact structural failure predictions, often leading to an overestimation of buildability.

Fluid-based computational approaches, on the other hand, allow for the geometrically-accurate simulation of the key stages of the 3D printing process, including pumping, extrusion, and layer deposition. Specifically, continuum fluid models developed using the Finite Volume Method (FVM) \cite{comminal2020, mollah2021, wolfs2021} and the Particle Finite Element Method (PFEM) \cite{reinold2022, rizzieri2023, rizzieri2025} have shown high accuracy in predicting filament shape and morphology. Similarly to continuum fluid approaches, meshfree methods have been applied to realistically simulate mixing and pumping in 3DCP \cite{ramyar2022, krenzer2022} or to estimate buildability \cite{zhu2023,chang2021}. However, the primary drawback of achieving high geometrical accuracy with continuum fluid or meshfree methods is their substantial computational cost, which remains a severe hindrance to their applicability in simulating full-scale structural elements \cite{rizzieri2024}.

To address this limitation, researchers have pursued two main strategies: improving the computational efficiency of numerical techniques, and developing simplified design tools derived from numerical models. Regarding the latter, a few studies have examined how material and printing parameters affect layer geometry, with the results presented in design charts for both rectangular \cite{liu2020} and circular nozzles \cite{rizzieri2025}. Nonetheless, while these tools provide fast and useful insights, their simplicity makes them suitable only for restricted cases, reducing their capacity to generalize.


\subsection{Data-driven approaches and machine learning in 3DCP}
As noted earlier, although numerical models can accurately predict and improve understanding of the printing process, their use remains constrained by high computational costs and complexity. Moreover, some tasks fall beyond their scope, either because the underlying physical relationships are not yet fully understood or because working directly with experimental data is more advantageous. This has driven growing interest in data-driven approaches and machine learning within the 3D printing community, with the goal of developing more advanced and accessible tools for digital construction \cite{geng2023}.

Machine learning is particularly effective for tasks where large datasets are available. In 3DCP, the automated nature of the process and the ability to collect extensive in-line data through continuously recording sensors \cite{wolfs2024} make data-driven approaches well-suited for quality control. For example, \cite{versteege2025} employed 2D laser scanning to capture the ‘as-printed’ layer geometry during production. The resulting geometric profiles were post-processed to extract relevant geometric parameters, commonly referred to as \textit{features}, including, e.g., layer dimensions, cross-sectional area, and surface texture, in order to reduce data complexity and facilitate the development of predictive models. Similarly, in \cite{cui2025} an AI-based computer vision method was applied to automatically extract the deformed layer’s contour and convert it into a set of features. The authors attempted to correlate layer deformation with rheological properties, though without being able to identify a clear trend.

While the above approaches mainly focused on data simplification and classification for an \textit{a posteriori} quality assessment, machine learning can also be harnessed for predictive purposes, such as estimating rheological or mechanical properties from mix composition. In \cite{charrier2022}, an Artificial Neural Network (ANN) was trained to predict mini-slump and dynamic yield stress as a function of the proportions between the components of the cement-based mixture. Likewise, \cite{gao2024} developed a data-driven model to predict plastic viscosity and yield stress of 3D-printable cementitious composites based on mixture composition and time after water addition. Predictions can also target final structural performance, as in \cite{schossler2025}, where multiple machine learning models, trained on a compilation of 3DCP mixtures from the literature and enhanced with Bayesian optimization, were used to forecast compressive strength, pump speed, and carbon footprint.

Finally, machine learning can also be leveraged to predict how material and printing parameters can affect the printing process, determining layer's shape.

In \cite{lao2020}, an ANN was trained on experimental data from a rectangular nozzle to predict cross-sectional shapes, simplified as flat-topped polygons with six control vertices per side, given the printing parameters and nozzle geometry. Similarly, in \cite{alhussain2024}, an \textit{ad hoc} experimental campaign produced a dataset used to train an ANN predicting filament width, contact width, and height from printing parameters. In contrast, \cite{silva2024} employed a synthetically generated numerical dataset to train an ANN for predicting filament widths starting from printing parameters, considering also generic non-rectilinear toolpaths and the possibility of partial material overlay.

While these studies achieved promising results and helped inspire the present work, they also reveal key limitations. First, the models were trained with fixed material parameters, requiring new datasets and re-training to adapt to different cementitious inks or environmental conditions. Second, experimental datasets rarely sample the parameter space uniformly, as reproducing all necessary material and printing conditions is often impractical within a single laboratory. Finally, the predicted outputs were predefined, oversimplified representations of the actual geometry, thus limiting the model’s flexibility to capture other user-relevant features. These constraints motivated the development of the deep learning tool for layer shape prediction presented in the following section.
 
\subsection{Proposed approach and plan of the article}
This work introduces a fast, accurate, and user-friendly deep learning framework for predicting filament shape geometry in 3D printing with cementitious materials, which has been named \textit{ShapeGen3DCP}. Unlike previous approaches, it integrates both material (density, yield stress and viscosity) and process parameters (nozzle diameter, nozzle height, printing velocity and flow velocity) as input variables and directly predicts the layer shape in terms of the precise cross-sectional profile. To mitigate experimental data scarcity, the training dataset was synthetically generated using a validated state-of-the-art numerical model of 3DCP \cite{rizzieri2023,rizzieri2025}. 

Before being fed to the neural network, some of the input variables are reformulated into physically meaningful dimensionless parameters. This reduces input dimensionality, improves robustness, and embeds prior physical knowledge by aggregating quantities known to be correlated. On the output side, geometries are compactly represented through Fourier descriptors, which expand the coordinates into a truncated Fourier series. This formulation inherently produces closed and smooth curves while reducing the prediction task to a small set of coefficients that can also enforce desirable properties such as symmetry. Overall, this representation simplifies the learning problem and accelerates training convergence.

The performance of the approach was assessed by comparing the generated cross-sections with both numerically simulated and experimentally tested cases not included in the training process. The results were excellent, demonstrating that the method is reliable and ready for real-world applications. For example, designers can use it to refine their products in advance, adjusting the toolpath to match predicted layer dimensions or enabling more complex designs with variable filament widths for optimized structures \cite{breseghello2022}. At the same time, the tool can assist practitioners in the field to reduce the need for trial-and-error tests, pre-calibrating printing settings to achieve the required layer shape.

The paper is organized as follows. Section \ref{sec:tool} introduces the machine learning tool for layer shape prediction from the user’s perspective. Section \ref{sec:dataset} details the dataset generation process, including the numerical 3DCP model used to produce synthetic data, the selected input parameters, and their variation ranges. Section \ref{sec:model} presents the deep learning model, beginning with the curves parametrization through Fourier descriptors, followed by a description of the neural network architecture and the training procedure. Section \ref{sec:results} first reports the training results of the proposed architecture together with a sensitivity analysis of its principal hyperparameters. Then, prediction results for new, unseen cases are compared with both numerical and experimental benchmarks. Section \ref{sec:discussion} examines the applicability range of the framework and suggests checks for filtering out cases of filament tearing or buckling. Finally, Section \ref{sec:conclusions} summarizes the main contributions and outlines possibilities for further improvement of the framework.

\section{Overview of the tool}\label{sec:tool}
\begin{figure}[h]
    \centering
    \includegraphics[width=0.65\linewidth]{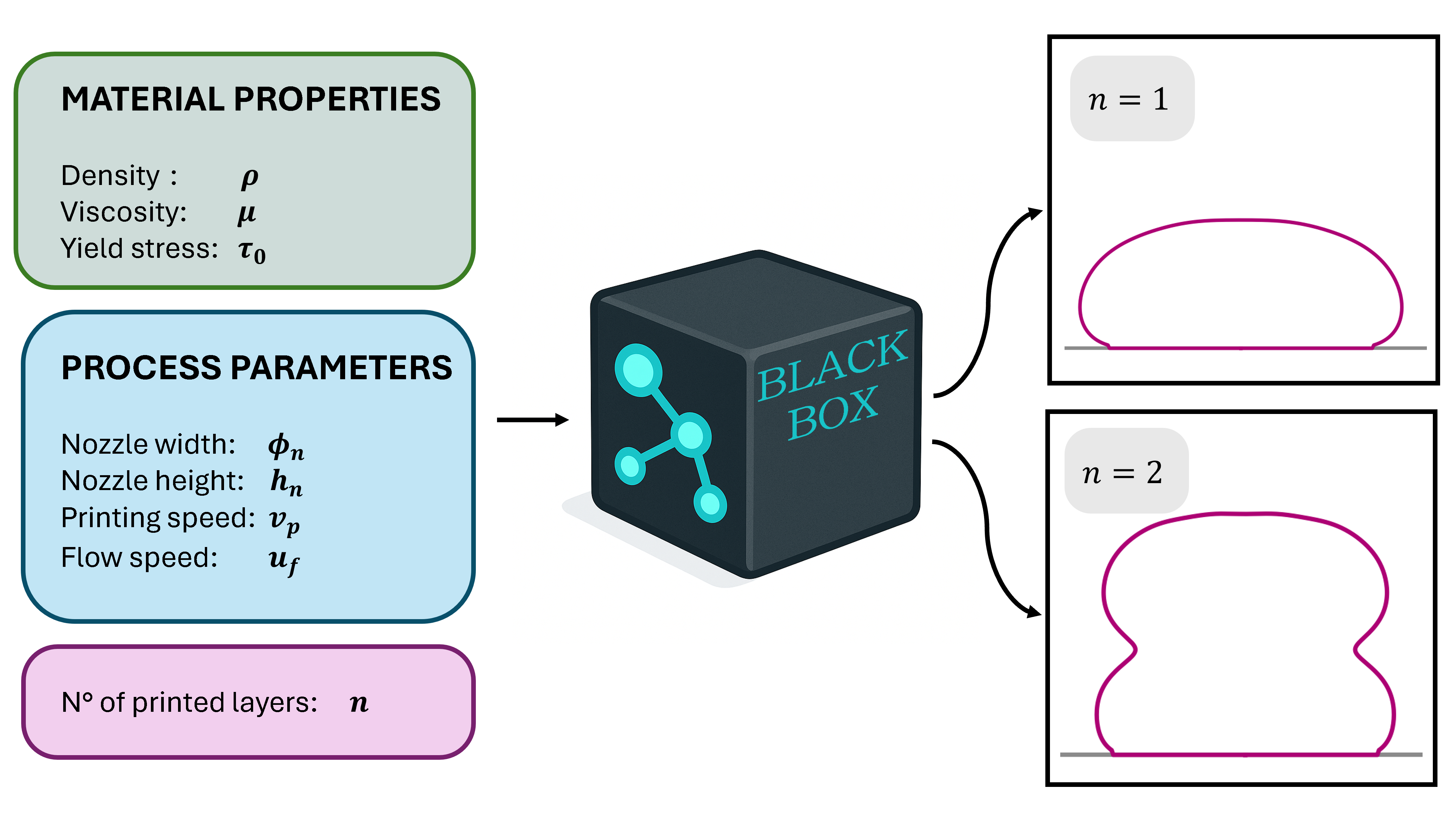}
    \caption{Schematic representation of \textit{ShapeGen3DCP} predictive framework. Given input material properties and process parameters, the model predicts the cross-section of the deposited material for one or two printed layers. }\label{fig:1}
\end{figure}

Figure \ref{fig:1} schematically illustrates the proposed black-box machine learning prediction tool. Inputs consist of material parameters (density, yield stress, viscosity) and process parameters (nozzle diameter, nozzle height, print velocity, flow velocity). The corresponding output is a set of Fourier descriptors that parameterize the predicted cross-sectional profile of the deposited filament.

The tool supports the prediction of both single-layer and two-layer configurations. An automated post-processing procedure extracts relevant geometrical parameters, hereafter referred to as \textit{features},from the output profiles, including layer width, height, cross-sectional area, and, in the case of two-layer walls, the inter-layer contact length.

While the tool's applicability is subject to the input parameter ranges defined in Section \ref{sec:dataset}, the use of dimensionless groups extends its effective coverage. Guidelines for detecting potential pathological cases (e.g. filament tearing, filament buckling) will also be presented in Section \ref{sec:discussion}.

An open-source, web-based implementation of \textit{ShapeGen3DCP} has been developed for public use and is available at the following page: \url{https://www.dica.polimi.it/ai3dcp}.

\section{Development of the dataset}\label{sec:dataset}
The dataset was generated from about 180 virtual printing simulations. Numerical simulations provide substantial advantages over experimental printing campaigns. Although adjusting individual printing parameters is feasible in practice, exploring a wide range of process parameter combinations alongside varying material properties becomes considerably more challenging. In experimental settings, this would typically require the design and preparation of new, tailor-made material mixes for each variation, making the process both time-consuming and costly.  

In the following, a brief description of the numerical model employed in this work is provided.  

\subsection{Numerical model}\label{subsec:numerical}
The numerical framework employed to perform the simulations and generate the dataset was first introduced in \cite{rizzieri2023}, where it was validated against experimental data for both single-layer and multi-layer printing scenarios under various combinations of printing parameters. It was subsequently validated against independent experimental data from other research groups, including cases involving different material formulations \cite{rizzieri2025}. This latter study also enabled an assessment of the influence of material and process parameters on the cross-sectional geometry of the deposited filaments. The model was later extended to incorporate time-dependent material behaviour \cite{rizzieri2023a} and to investigate instability phenomena during printing \cite{rizzieri2024}.

In this work, the model is applied to generate multiple virtual prints involving the extrusion of two rectilinear layers. For completeness, the fundamentals of the model and its main underlying assumption are briefly discussed below.

\subsubsection{Governing equations}
Fresh concrete is assumed to behave as a homogeneous continuum fluid. Thus, its behavior is governed by the Navier–Stokes equations, accounting for the balance of momentum and mass:

\begin{align}
\rho \frac{d \bm{u}}{d t} = \nabla_{\bm{x}} \cdot \bm{\sigma} + \rho \bm{b} \quad \text{in } \Omega_{t} \times [0, T], \label{eq:1} \\
\ \nabla_{\bm{x}} \cdot \bm{u} = 0 \quad \text{in } \Omega_{t} \times [0, T], \label{eq:2}
\end{align}
\noindent where $\rho$ is density assumed to be constant, $\bm{u}=\bm{u}(\bm{x},t)$ is the velocity field, $p=p(t)$ is the pressure field, $\bm{\sigma}= \bm{\sigma}(\bm{x},t)$ is the Cauchy stress tensor, $\bm{b}$ is the vector of the external accelerations and $\frac{d (\bullet)}{dt} = \frac{\partial (\bullet)}{\partial t}\Big\vert_{\bm{x}} + \bm{c} \cdot \nabla_{\bm{x}}(\bullet)$ represents the total time derivative, with $\bm{c}$ being the convective velocity.

Additionally, it is possible to decompose the 
 Cauchy stress tensor in a volumetric and a deviatoric part:
\begin{equation}
    \bm{\sigma} = -p \bm{I} + \bm{\tau},   \label{eq:3}
\end{equation}
\noindent where $\bm{I}$ is the identity tensor and $\bm{\tau}$ is the deviatoric stress tensor. The deviatoric stress tensor can be computed after the definition of an adequate rheological law. For fresh concrete, the Bingham model is often adopted, which accounts for the presence of a yield stress in the material: if the stresses are lower than the yield stress, the fluid behaves as a rigid body, while if the yield stress threshold is overcome, the fluid flows. The Bingham model reads:

\begin{align}
\bm{\tau}(\bm{u})&=2\mu\bm{\epsilon}(\bm{u})+\tau_0 \frac{\bm{\epsilon}(\bm{u})}{\|\bm{\epsilon}(\bm{u})\|} \quad &\text{if }\|\bm{\tau}\|>\tau_0, \label{eq:4} \\
\bm{\epsilon}(\bm{u})&=0 \quad &\text{otherwise}, \label{eq:5} 
\end{align}

\noindent where $\mu$ is the Newtonian viscosity, $\tau_0$ is the yield stress, $\| \bullet \|$ is the Von Mises norm and  $\bm{\epsilon}(\bm{u})$ is the deviatoric strain rate tensor defined, for an incompressible fluid, as $\bm{\epsilon}(\bm{u})=  \frac{1}{2} (\nabla_{\bm{x}} \bm{u}+ \nabla_{\bm{x}} \bm{u}^T )$.

Due to the discontinuous nature of the Bingham law and the sharp fluid-to-solid transition, which may lead to numerical instabilities, the model relies on the implementation based on an exponential regularization \cite{papanastasiou1987}.

It should be noted that Bingham’s model neglects both elasticity and shear-thinning effects. Time-dependent variations of its parameters are likewise disregarded, as thixotropy and hydration phenomena generally manifest on time scales much longer than those relevant to extrusion and the deposition of a few layers. Surface tension is also omitted; as discussed in \cite{roussel2018}, this assumption remains valid when the characteristic extrusion dimensions exceed a few millimetres.

\subsubsection{Space and time discretizations}
In this work, the Navier–Stokes equations are solved using the Finite Element Method (FEM). The continuum domain is discretized with a finite element mesh employing standard Galerkin shape functions. For reasons that will be clarified later, linear shape functions are used, i.e., triangular elements in 2D and tetrahedral elements in 3D. It is worth noting that this choice, which employs shape functions of the same order for both velocity and pressure fields, violates the Ladyzhenskaya–Babuška–Brezzi (LBB) stability condition. To address this limitation and allow the safe use of equal-order velocity-pressure pairs, the Pressure Stabilizing Petrov-Galerkin (PSPG) method \cite{Hughes1986} is employed. Time discretization is then performed by subdividing the time interval of interest into finite time steps $\Delta t$ and approximating derivatives using the implicit backward Euler scheme. Further details regarding the spatial and temporal discretization, as well as the solution of the resulting linearized algebraic system, can be found in \cite{rizzieri2025a}.

\subsubsection{Particle Finite Element Method (PFEM)}
In this numerical model, the governing equations are written and solved using a Lagrangian framework across most of the computational domain. This approach offers several advantages, including the inherent ability to capture free surfaces and the elimination of nonlinear convective terms in the equations of motion. However, in problems involving large deformations or displacements, the Lagrangian approach will cause a deterioration of the computational mesh, causing accuracy and stability issues.

An appealing remedy is given by the so-called ``Particle Finite Element Method (PFEM)'' \cite{onate2004,cremonesi2020}, which equips the Lagrangian framework with an efficient re-meshing technique. The PFEM was originally developed for the simulation of free surface flows and breaking wave problems \cite{Idelsohn2004}. Nonetheless, due to its versatility and robustness PFEM was soon applied in many other engineering fields \cite{cremonesi2020}, among which also complex fluids \cite{rizzieri2024a,rizzieri2025a} and additive manufacturing applications \cite{reinold2022,reinold2024,rizzieri2023,rizzieri2025,rizzieri2024}.

\subsubsection{3D printing PFEM framework}
As previously mentioned, PFEM provides a versatile and robust framework particularly well-suited for modeling free-surface flows under complex and time-varying boundary conditions. For these reasons, it appears an ideal candidate for the accurate simulation of the 3D printing process. In particular, the framework must be complemented with a set of \textit{ad hoc} techniques designed to efficiently reproduce the extrusion and layer deposition processes. These techniques were developed and described in detail in \cite{rizzieri2023, rizzieri2025}; here, only the main challenges encountered are outlined, together with an overview of the approaches adopted to address them.

\begin{itemize}
    \item \textbf{Nozzle movement and continuous material flow} \\
    To simulate the rigid motion of the nozzle in 3D printing according to $G$-code instructions, the ALE framework \cite{donea2017} is employed. By decoupling mesh and material motion, the application of multiple boundary conditions at the nozzle outlet is facilitated. Specifically, the printing velocity at the nozzle, $\bm{v} = \bm{v}_{p}$, is prescribed as the mesh velocity, while the inflow velocity, $\bm{u} = \bm{u}_f$, is imposed as a standard Dirichlet boundary condition. Additionally, to manage element stretching below the nozzle caused by the inflow, new nodes are inserted and re-meshing is performed in the computational domain whenever necessary, ensuring continuous material flow. 

    \item \textbf{Contact between the filament and the rigid print bed} \\
Generally, in PFEM, perfectly rigid constraints are pre-materialized by inserting fixed nodes along the Dirichlet boundary \cite{cremonesi2020}. This approach can cause non-negligible volume variations, especially on coarse meshes. For this reason, a novel contact strategy has been proposed in \cite{rizzieri2023} to accurately capture the interaction between the extruded filament and the print bed. Specifically, Dirichlet Boundary Conditions (DBCs) are imposed by “freezing” (setting the velocity to zero) those nodes which satisfy a prescribed positional constraint (e.g., $z \leq 0$) to replicate the effect of the contact with the plane.
\end{itemize}

\subsection{Dataset generation}\label{subsec:dataset}
\begin{figure}[h!]
    \centering
    \includegraphics[width=0.75\linewidth]{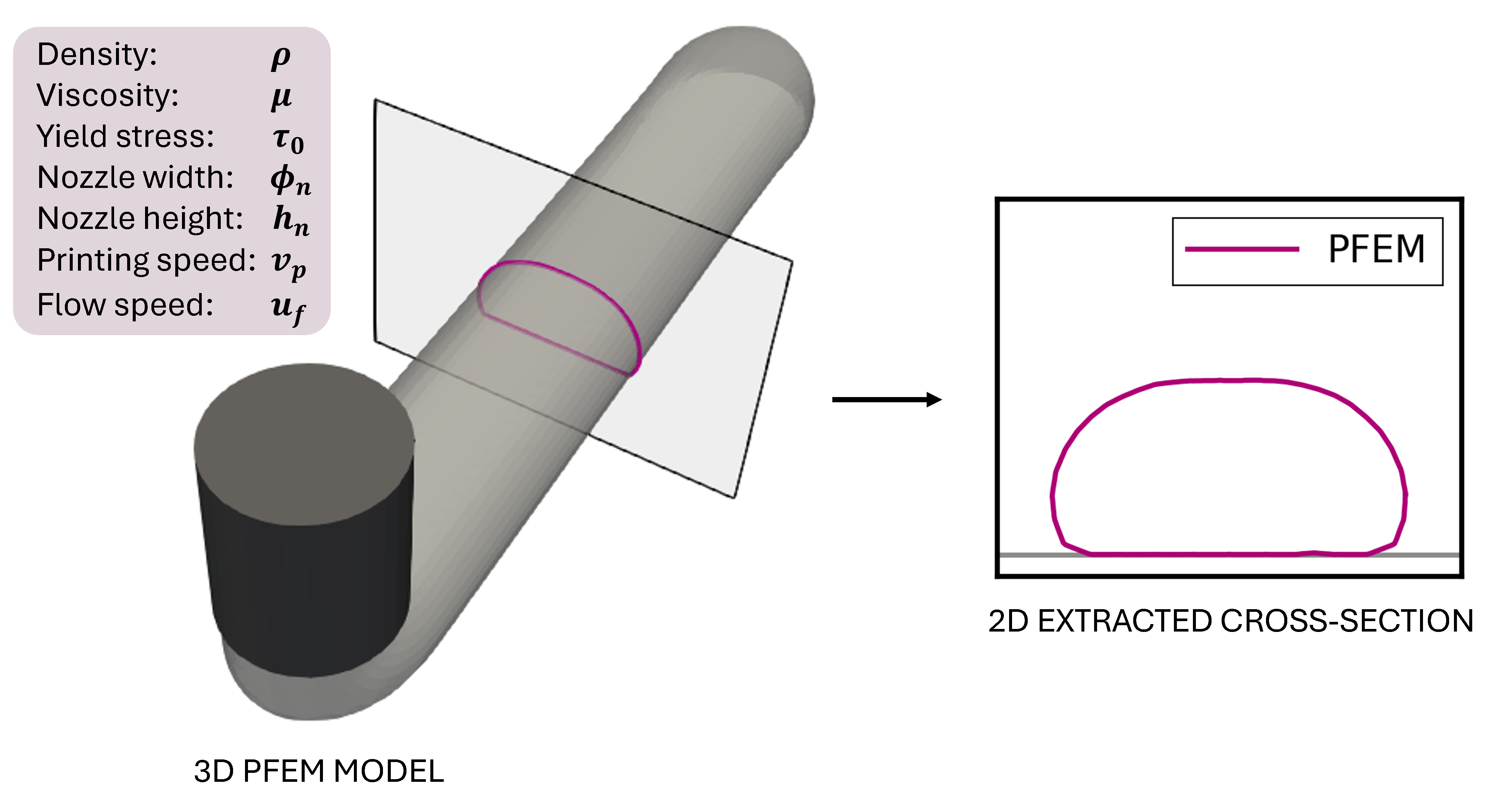}
    \caption{Numerical computation of the 3D filament geometry using the PFEM model, followed by extraction of the 2D midspan cross-section.
}\label{fig:2}
\end{figure}

A well-structured dataset is crucial for effectively training machine learning models and ensuring reliable predictive performance. In this study, the previously described numerical model was used to generate two synthetic datasets, of about 180 and 170 unique printing scenarios, respectively. The first dataset is intended for predicting single-layer scenarios, while the second targets two-layer cases. Only printing cases involving circular nozzles are considered. The printing is carried out on a rigid planar support (i.e., the print bed) and a rectilinear toolpath is adopted. Figure~\ref{fig:2} illustrates the methodology used to extract cross-sections for a representative case.

Each printing scenario is defined by a unique combination of seven input parameters ($\rho, \ \tau_0, \ \mu, \ \phi_n, \ h_n, \ v_p, \ u_f$), covering both material properties and process variables. The physically meaningful limit ranges for each parameter are reported in Table~\ref{tab:1}. They were selected based on relevant literature (e.g., \cite{wolfs2021}) and the authors’ expertise in order to ensure broad applicability.

\begin{table}[h]
\centering
\caption{Definition of the seven dimensional input parameters with their corresponding ranges of variation.}
\label{tab:1}
\begin{tabular*}{0.9\linewidth}{@{\extracolsep{\fill}} l l l l l }
\toprule
 & \textbf{Input parameter} & \textbf{Symbol} & \textbf{Range} & \textbf{Unit} \\
\midrule
\multirow{3}{*}{\textbf{Material parameters}} 
    & Density        & $\rho$   & 2000 -- 2500       & $\mathrm{kg\,m^{-3}}$ \\
    & Viscosity      & $\mu$    & 1 -- 30            & $\mathrm{Pa \cdot s}$ \\
    & Yield stress   & $\tau_0$ & 100 -- 1500        & $\mathrm{Pa}$ \\
\addlinespace
\multirow{4}{*}{\textbf{Process parameters}} 
    & Nozzle diameter & $\phi_n$ & 5 -- 30            & $\mathrm{mm}$ \\
    & Nozzle height   & $h_n$    & 5 -- 30            & $\mathrm{mm}$ \\
    & Printing speed  & $v_p$    & 10 -- 300          & $\mathrm{mm\,s^{-1}}$ \\
    & Extrusion speed & $u_f$    & 10 -- 300          & $\mathrm{mm\,s^{-1}}$ \\
\bottomrule
\end{tabular*}
\end{table}

The design of the dataset is critical, as machine learning models can only generalize accurately if the training data sufficiently explores the input parameter space. Poor sampling strategies that over-represent certain regions while neglecting others can produce models that are prone to overfitting or that fail to capture key physical relationships. Therefore, it is essential to cover the high-dimensional parameter space as uniformly as possible. To this end, Latin Hypercube Sampling (LHS) \cite{mckay1979} was employed, a sampling strategy designed to evenly cover the parameter space, effectively stratifying the sampling across all marginal distributions and minimizing clustering.

The LHS method was applied to sample uniformly within the defined parameter bounds. The resulting combinations of input parameters were used to configure numerical simulations of rectilinear prints comprising two layers. Some input parameter combinations led to failed depositions, for instance, due to filament tearing, buckling, or slug formation, and were therefore identified and excluded from the dataset. For the remaining valid cases, the filament length was adjusted to ensure a stationary printing regime was achieved. To minimize boundary effects, cross-sections were extracted at the mid-span of the printed structure for both the first and second layers. Consequently, each entry in the final dataset corresponds to a unique combination of the seven input parameters and the associated cross-sectional geometry.

As mentioned  before, two separate datasets were generated: one for predicting single-layer cross-sections and another for two-layer cross-sections. Following standard machine learning practice, each dataset was split into training, validation, and test sets. For the single-layer dataset, the split was 154:14:16 samples, while for the two-layer dataset it was 154:12:6. The test sets were specifically designed to include only cases in which experimental data are also available, enabling direct validation of the model’s predictive performance against physical benchmarks.

\subsubsection{Use of dimensionless quantities}
Notably, as demonstrated in recent studies \cite{carneau2022,wolfs2021,rizzieri2025}, the effects of material and process parameters are often more effectively characterized by dimensionless groups rather than by examining individual parameters. To inject this prior physical knowledge directly into the deep learning model, the following dimensionless quantities are introduced:
\begin{align}
 \tau_0^* &= \frac{\tau_0}{\rho g \phi_n}, \label{eq:12}\\
 v^* &= \frac{v_p}{u_f},
 \label{eq:13}
\end{align}
\noindent where $\tau_0^*$ is the dimensionless yield stress, condensing the information regarding the yield stress, the density and the nozzle diameter. It represents the capacity of the material’s microstructure to balance gravity-induced stresses. The dimensionless velocity ratio $v^*$ incorporates instead the printing and extrusion velocities and describes the flow regime generated during the extrusion process. This aggregation of correlated raw parameters into physically meaningful groups effectively reduces the input dimensionality and teaches the model the fundamental relationships between variables \textit{a priori}, rather than requiring it to learn these complex correlations from the data alone.

Consequently, the model's input space is reduced to two dimensionless parameters ($\tau_0^*$, $v^*$) and three dimensional ones ($\mu$, $\phi_n$, $h_n$). This enables the model to capture physical dependencies more straightforwardly than if it were trained solely on raw inputs. Furthermore, this approach enhances extrapolation capability: even if some raw parameters fall outside the training range, the model can still provide accurate predictions as long as they appear through the dimensionless groups and those groups remain within the learned domain. Consequently, the model improves generalization to unseen cases while maintaining physical consistency. The ranges of this reduced input set, which define the effective applicability domain of the model, are reported in Table \ref{tab:2}.


\begin{table}[h]
\centering
\caption{Definition of the reduced input set, consisting of both dimensional and dimensionless parameters and their limit ranges of variation.}
\label{tab:2}
\begin{tabular*}{0.9\linewidth}{@{\extracolsep{\fill}} l l l l l }
\toprule
 & \textbf{Input parameter} & \textbf{Symbol} & \textbf{Selected range} & \textbf{Unit} \\
\midrule
\multirow{2}{*}{\textbf{Material parameters}} 
    & Dimensionless yield stress & $\tau_0^*$ & $\sim 0.1$ -- $7.6$ & -- \\
    & Viscosity & $\mu$ & 1 -- 30 & $\mathrm{Pa \cdot s}$ \\
\addlinespace
\multirow{3}{*}{\textbf{Process parameters}} 
    & Nozzle diameter & $\phi_n$ & 5 -- 30 & $\mathrm{mm}$ \\
    & Nozzle height   & $h_n$ & 5 -- 30 & $\mathrm{mm}$ \\
    & Dimensionless velocity ratio & $v^*$ & $\sim 0.03$ -- $30$ & -- \\
\bottomrule
\end{tabular*}
\end{table}

\section{Deep learning framework}\label{sec:model}
In additive manufacturing, predicting the geometry of printed elements from process parameters has become increasingly important. While this task was once the domain of numerical models, deep learning approaches are now emerging as powerful alternatives. In deposition-based processes, a key challenge is defining a contour representation of the deposited material that balances two requirements: it should incorporate prior knowledge of the expected output while remaining expressive enough to capture geometric details, yet compact and efficient for neural network training.

One possible approach is to represent contours as unstructured point clouds. This strategy can achieve accurate reconstructions, as demonstrated in the context of Cold Spray Additive Manufacturing (CSAM) \cite{liu2021}. However, because the prediction lacks geometrical constraints, the resulting curves are often noisy and irregular. An alternative is to use a curve parameterization that embeds prior geometric knowledge. For instance, in \cite{ikeuchi2024}, the contours of the deposed material in CSAM, have been modelled as a superposition of Gaussian functions, inherently enforcing smoothness and symmetry.

In \cite{vanerio2025} a neural network was trained to predict cross-sections in Fused Granulate Fabrication (FGF). In this case accurate reconstructions of the shapes of the extruded material were obtained by modeling cross-sections as binary images. While this guarantees achieving a closed profile, it does not ensure smoothness nor symmetry.

To overcome these limitations, the present work introduces a deep learning framework based on a Fourier descriptor parametric curve representation of the contours, embedding geometric properties such as smoothness, closure, and symmetry directly into the model. The methodology is applied for predicting layer shapes obtained by extrusion of yield stress fluids, as in the case of 3D printing with cementitious materials.

\subsection{Curve parametrization with Fourier descriptors}\label{subsec:fourier}
Fourier descriptors \cite{zahn1972, kuhl1982} provide a compact and efficient representation of closed 2D contours by expanding the coordinates into a truncated Fourier series. The periodicity of the basis functions guarantees closure and continuity. Each harmonic has a clear geometric interpretation, corresponding to an ellipse defined by its coefficients and frequency. Therefore, the contour is reconstructed as a superposition of ellipses. This captures both global shape and local features with relatively few coefficients, making it a widely used method for representing arbitrary closed shapes in functional form \cite{yadav2007}. In computer vision, for example, Fourier descriptors have been used to represent droplet boundaries from experimental images \cite{durve2025}. In this work, Fourier descriptors are employed to represent 3DCP filament cross-sections, as they provide a compact and interpretable representation while inherently ensuring closure, smoothness, and symmetry of the predicted shapes.

Formally, let us consider a closed 2D contour $\Gamma(t)$ parameterized by $t \in [0, 2\pi]$, with coordinates $(x(t), y(t))$. In the Fourier descriptors framework, this contour can be represented by a truncated Fourier series expansion for each coordinate:

\begin{equation}
\left\{
\begin{array}{l}
x(t) = a_0^x + \sum\limits_{k=1}^{N-1} \left[ a_k^x \cos(k t) + b_k^x \sin(k t) \right] \\[1em]
y(t) = a_0^y + \sum\limits_{k=1}^{N-1} \left[ a_k^y \cos(k t) + b_k^y \sin(k t) \right]
\end{array}
\right.
\quad \text{with } k = 1, \dots, N-1,
\end{equation}

where $N$ is the total number of harmonics. The Fourier coefficients for each coordinate are defined as:

\begin{align}
\begin{aligned}
a_0^x &= \frac{1}{2\pi} \int_0^{2\pi} x(t) \, dt, &
a_0^y &= \frac{1}{2\pi} \int_0^{2\pi} y(t) \, dt, \\
a_k^x &= \frac{1}{\pi} \int_0^{2\pi} x(t) \cos(k t) \, dt, &
a_k^y &= \frac{1}{\pi} \int_0^{2\pi} y(t) \cos(k t) \, dt, \\
b_k^x &= \frac{1}{\pi} \int_0^{2\pi} x(t) \sin(k t) \, dt, &
b_k^y &= \frac{1}{\pi} \int_0^{2\pi} y(t) \sin(k t) \, dt.
\end{aligned}
\end{align}

The constant terms $a_0^x$ and $a_0^y$ define the centroid of the shape, while the other represent the shape’s contour information, capturing its size, orientation, and detailed geometric features at different frequencies.

To enforce symmetry with respect to the $y$-axis, Fourier coefficients are constrained according to the parity requirements of Fourier basis functions. Indeed, the $x$-coordinate must be an odd function, satisfying $x(-t) = -x(t)$, while the $y$-coordinate must be an even function, satisfying $y(-t) = y(t)$. Given the parity properties of the basis functions, where cosine terms $\cos(kt)$ are even and sine terms $\sin(kt)$ are odd, the $x$-coordinate must be represented exclusively by sine terms, and the $y$-coordinate exclusively by cosine terms. This ensures that for every point $(x(t), y(t))$ always exists a mirror point represented in the parametrization $(-x(t), y(t))$, thereby guaranteeing the desired symmetry. Practically, this is achieved by imposing:
\begin{equation}
a_k^y = 0, \ b_k^x = 0 \quad \forall \ k = 1, \dots, N-1 \quad \text{and} \quad a_0^x = 0. 
\end{equation}

Thus, the symmetric contour is fully parameterized by a vector $\bm{f} = (a_0^y, a_1^x, \dots, a_{N-1}^x, b_1^y, b_2^y, \dots, b_{N-1}^y) \in \mathbb{R}^{\widetilde{N}}$ of coefficients, where $\widetilde{N} = 2N - 1$.

\begin{figure}[h]
    \centering
    \subfloat[$\widetilde{N}=9$]{\includegraphics[width=0.333\linewidth]{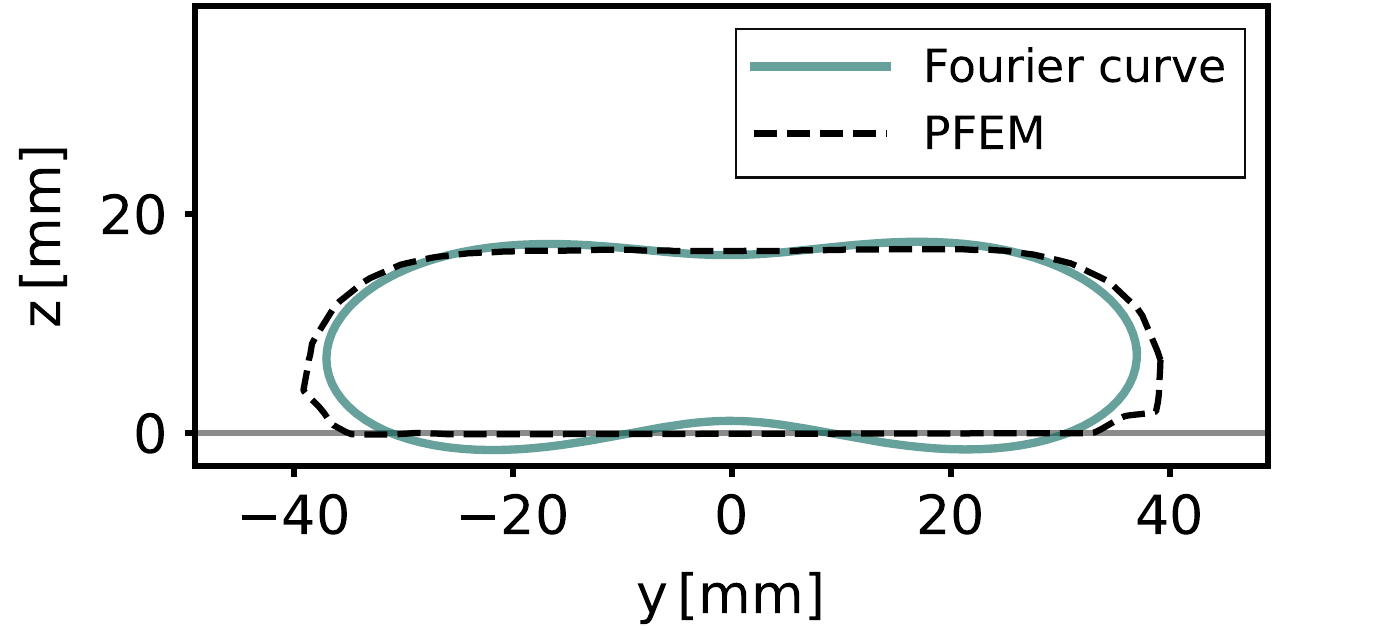}} 
    \subfloat[$\widetilde{N}=21$]{\includegraphics[width=0.333\linewidth]{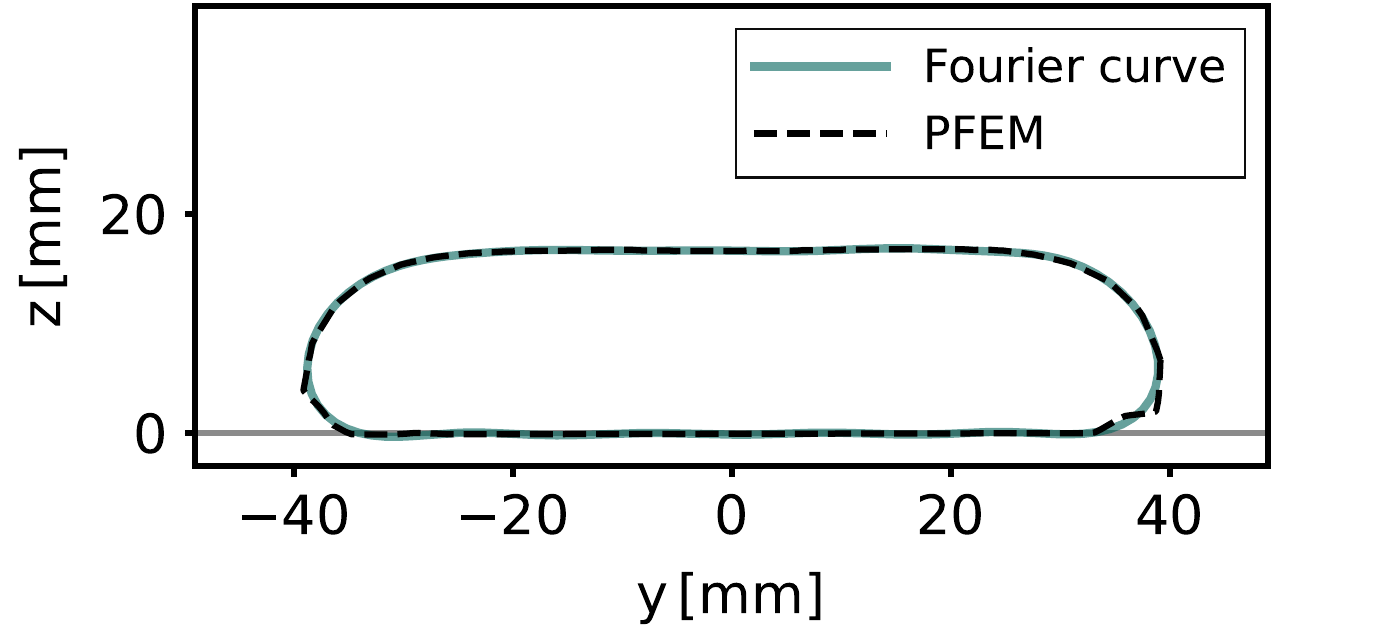}}
    \subfloat[$\widetilde{N}=39$]{\includegraphics[width=0.333\linewidth]{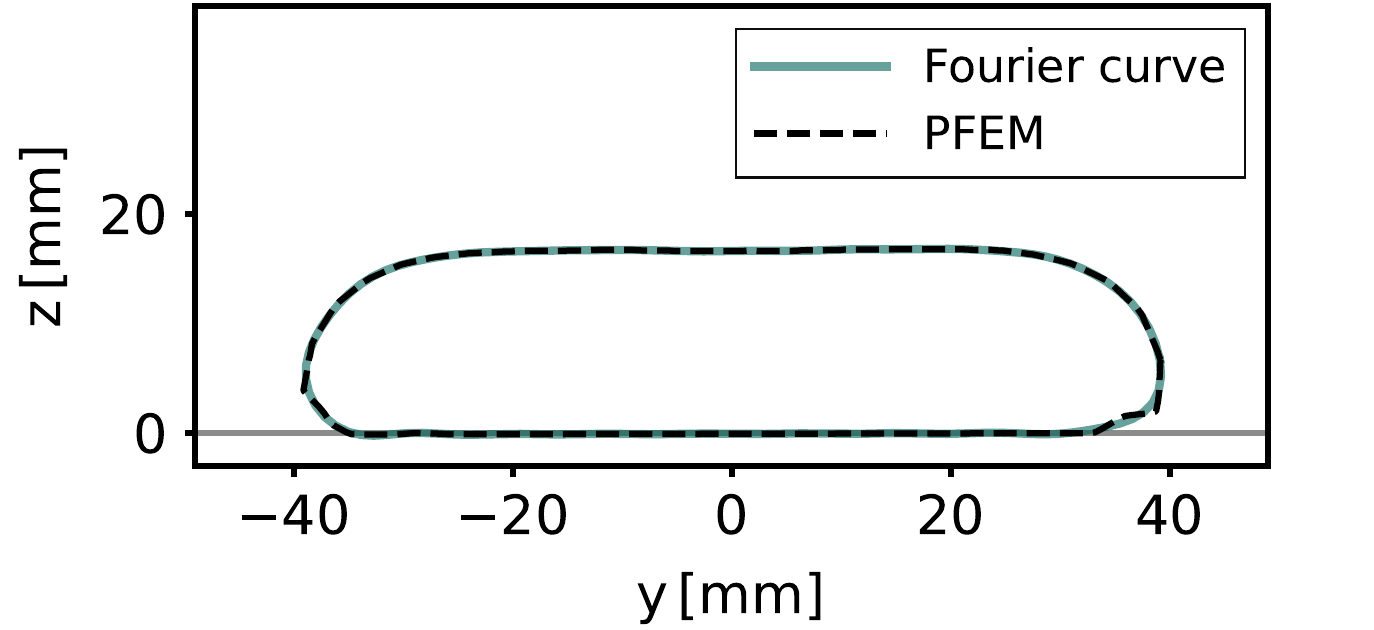}}\\
    \subfloat[$\widetilde{N}=9$]{\includegraphics[width=0.333\linewidth]{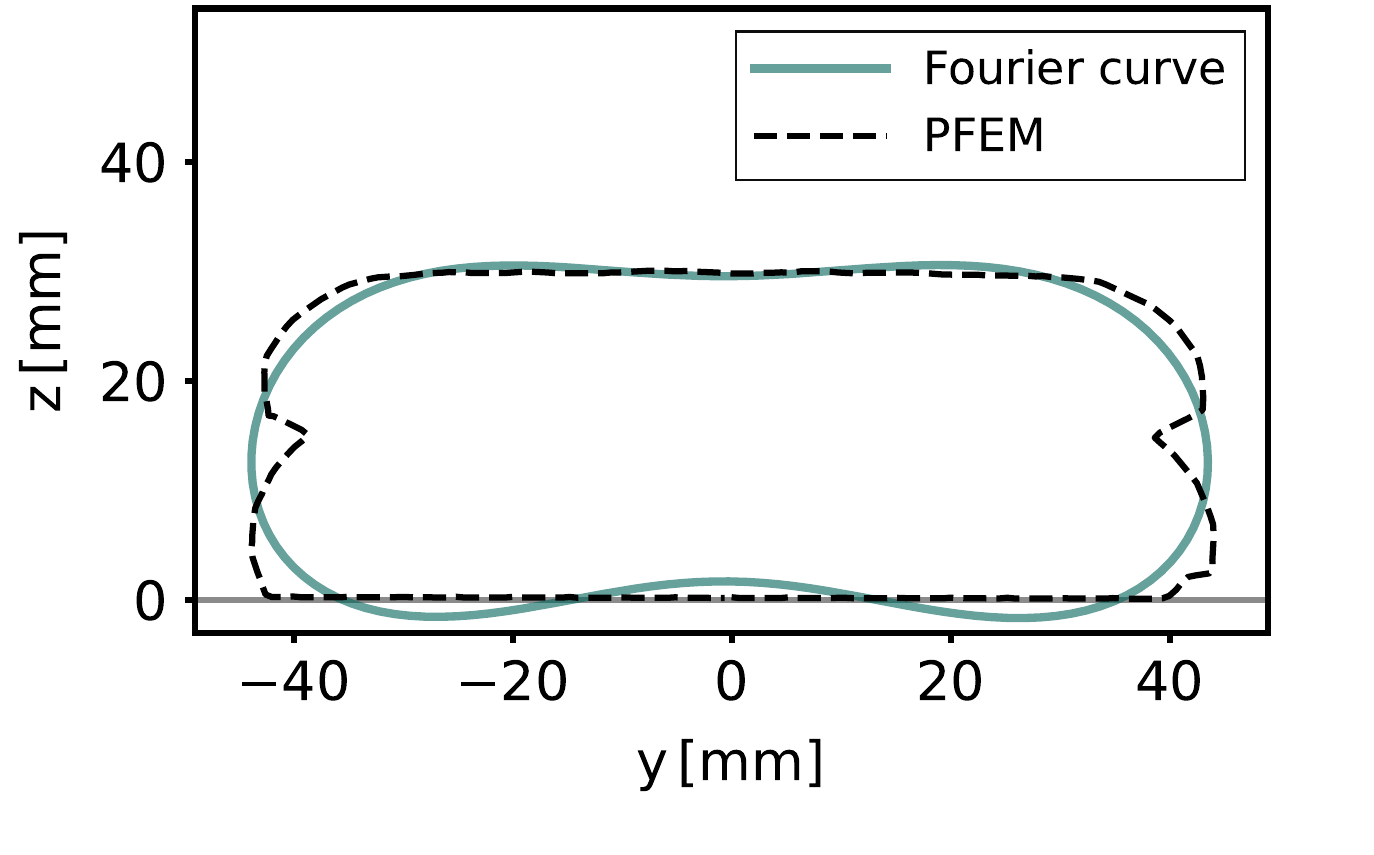}}
    \subfloat[$\widetilde{N}=21$]{\includegraphics[width=0.333\linewidth]{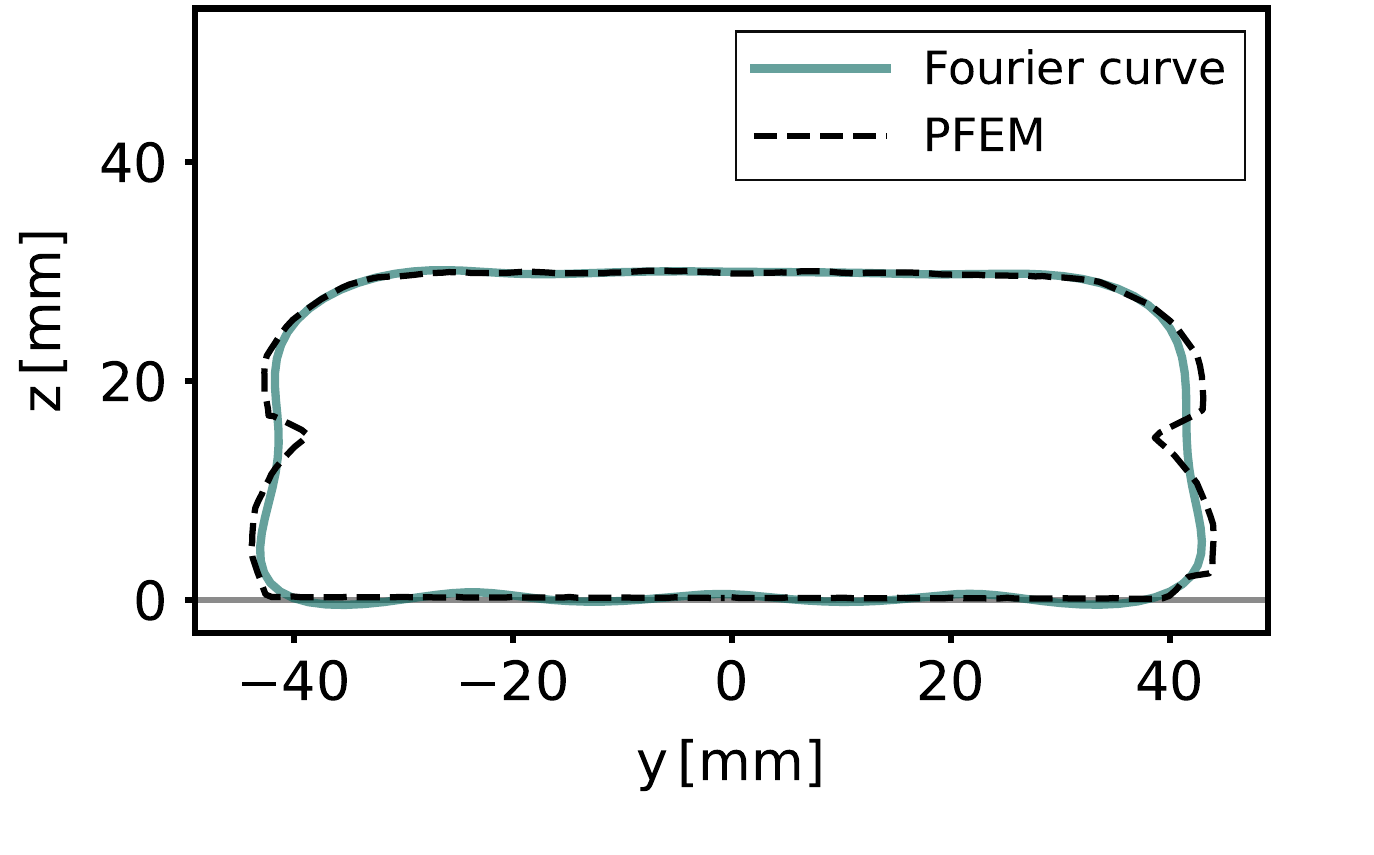}}
    \subfloat[$\widetilde{N}=39$]{\includegraphics[width=0.333\linewidth]{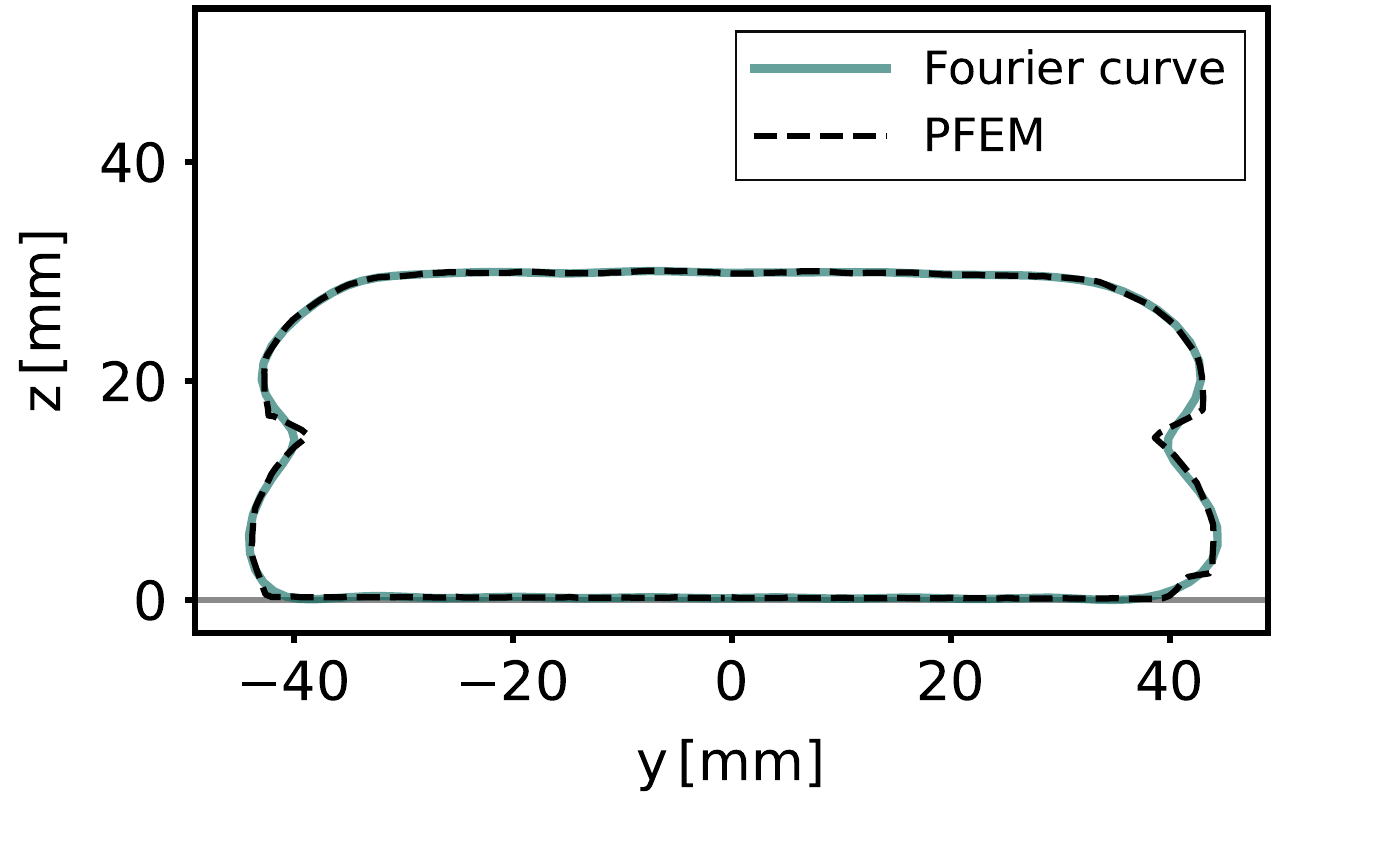}}
    \caption{Fourier descriptors approximations of filament cross-sections varying the number of coefficients. As the number of harmonics increases, the reconstructed contours progressively approach the true shape.}
    \label{fig:3}
\end{figure}

Figure \ref{fig:3} illustrates how the reconstruction accuracy varies as the number of harmonics increases. A low number of coefficients captures the overall shape, but lacks detail. The reconstruction converges to the true contour as more coefficients are added. Cross-sections involving two layers present more complex shapes, characterized by sharp angles in correspondence of the contact region, and consequently require a higher number of Fourier coefficients for an accurate approximation. The number of Fourier descriptors $N$ is a fundamental model hyperparameter; its selection is discussed in Section \ref{sec:train_res}.

Once the Fourier parametrization has been introduced to represent 3DCP cross-sectional contours, it becomes possible to design a deep learning framework that learns the corresponding Fourier coefficients. Specifically, the model learns a mapping $\mathcal{M}$ from the set of $K$ input parameters $\bm{p} \in \mathbb{R}^K$ to the vector $\bm{f}\in \mathbb{R}^{\widetilde{N}}$ of non-zero Fourier descriptor coefficients that define the cross-section contour:

\begin{equation}
\begin{aligned}
\mathcal{M}: \mathbb{R}^K & \to \mathbb{R}^{\widetilde{N}} \\
\bm{p} & \mapsto \bm{f} = \mathcal{M}(\bm{p}).
\end{aligned}    
\end{equation}
In the following, the input parameter vector is defined as 
$\bm{p} = (\tau_0^*, \mu, \phi_m, h_n, v^*)$, 
which includes both material and process parameters, for a total of $K = 5$. Before training, the five input parameters are normalized to the range [0,1] based on statistics computed on the dataset to ensure that all input features are on the same scale, a standard practice that helps convergence and stabilize the training process. 

\subsection{Neural network architecture}\label{subsec:network}

 \begin{figure}[h!]
    \centering
    \includegraphics[width=0.76\linewidth]{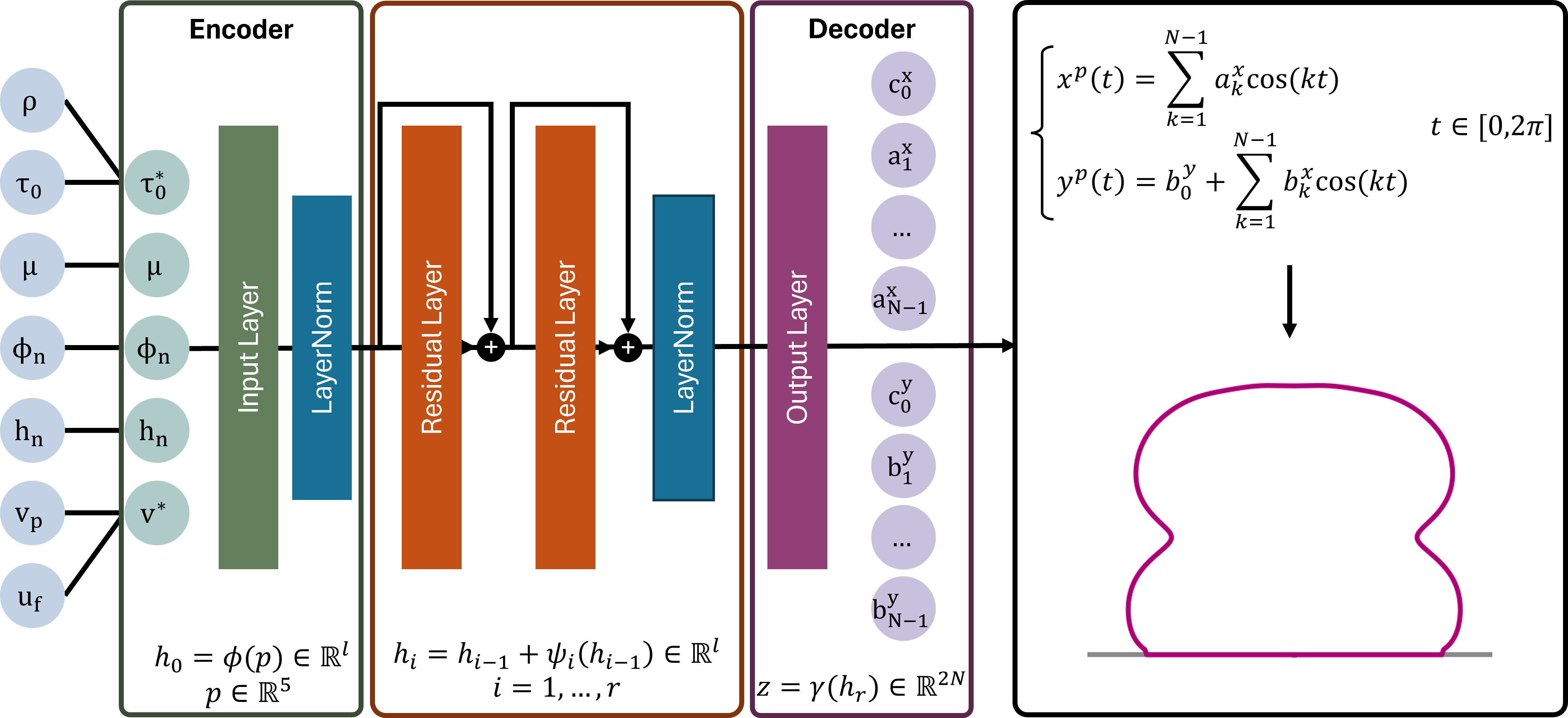}
    \caption{\textit{ShapeGen3DCP} architecture, consisting of three blocks: an encoder that maps input parameters into a latent space, a processor with $r$ residual layers, and a decoder projecting back to the output space. The predicted Fourier coefficients are then used to reconstruct the closed curve representing the filament cross-sectional profile.}\label{fig:4}
\end{figure}

The neural network architecture of \textit{ShapeGen3DCP} is schematized in Figure \ref{fig:4}. The model firstly projects the input vector $\bm{p}\in\mathbb{R}^5$ into a higher-dimensional feature space through the transformation:
\begin{equation}
    \bm{h}_0 = \bm{\phi}(\bm{x}), \quad \bm{h}_0\in \mathbb{R}^l, 
\end{equation}
where $l$ represents the feature dimension and $\bm{\phi}$ an encoder function. The encoded features are then processed through a residual block, constituted by $r$ sequential layers $\bm{\psi_i}, \, i=1,\ldots,r$, each layer incorporating skip connections:
\begin{equation}
    \bm{h}_i = \bm{h}_{i-1} + \bm{\psi}_i(\bm{h}_{i-1}), \quad \bm{h}_i \in \mathbb{R}^l, \, i=1,\ldots,r.
\end{equation}
Finally, the processed features $\bm{h_r}$ are mapped to the output space through a decoder function $\bm{\gamma}$:
\begin{equation}
   \bm{f} = \bm{\gamma}(\bm{h}_r), \quad \bm{h} \in \mathbb{R}^{\widetilde{N}}, 
\end{equation}
that produces the output vector $\bm{f}$ of the predicted Fourier coefficients.
These coefficients are associated with a continuous representation of the cross-sectional curve, from which a variable number $N_p$ of points $\bm{x}_i^p = [x_i^p, y_i^p], \, i = 1, \ldots, N_p$ can be flexibly sampled at equidistant values of $t \in [0,2\pi]$.

Throughout the network, normalization layers are integrated to improve convergence properties and stabilize training, while residual connections allow gradients to propagate more effectively through the network by providing shortcut paths that bypass layers. This enables the training of deeper architectures without loss of information. $\tanh$ is employed as an activation function. 

\subsection{Training procedure}\label{subsec:training}
The encoder function $\phi$, the processor functions $\psi_i$, and the decoder function $\gamma$ are learnable functions, each associated with a set of learnable weights. These weights are optimized during training using the AdamW optimizer \cite{loshchilov2019}, a variant of stochastic gradient descent. The network is trained to minimize a loss function composed of two components: a point-wise reconstruction loss and a smoothness regularization loss. The reconstruction loss is defined as the mean Euclidean distance between the predicted curve points $\bm{x}_i^p$ and the corresponding ground-truth points $\bm{x}_i$, normalized by the maximum norm of the ground-truth points to ensure scale invariance. The smoothness term penalizes instead discrepancies between the derivatives of the predicted and ground-truth, $\bm{x}_i^{p\prime}$ and $\bm{x}_i^\prime$ respectively, with respect to the curve parameter $t$ and it is also normalized by the maximum derivative magnitude. The relative weight of the smoothness term is controlled by the hyperparameter $\lambda$. The total loss is given by:

\begin{equation}
\mathcal{L} = \frac{1}{N_p} \sum_{i=1}^{N_p} \frac{\left\| \bm{x}_i^p - \bm{x}_i \right\|_2}{\max_j \left\| \bm{x}_j \right\|_2}
+ \lambda \cdot \frac{1}{N_p - 2} \sum_{i=2}^{N_p - 1} \frac{\left\| \bm{x}_i^{p\prime} - \bm{x}_i^\prime \right\|_2}{\max_j \left\| \bm{x}_j^\prime \right\|_2},
\label{eq:loss_normalized}
\end{equation}
where the derivatives with respect to the parameter $t$ are approximated using centered finite differences as:

\begin{equation}
\mathbf{x}_i^\prime \approx \frac{\bm{x}_{i+1} - \bm{x}_{i-1}}{t_{i+1} - t_{i-1}}, \quad i = 2, \ldots, N_p - 1.
\label{eq:finite_difference}
\end{equation}

The learning rate, which determines the step size for updating model weights at each iteration of gradient descent, is initially set to 0.001. To improve convergence, the learning rate is gradually reduced over epochs. This strategy allows the model to make larger updates early in training, promoting effective exploration of the parameter space, and then progressively refines the updates as training progresses, enabling more precise convergence toward an optimal solution. Training is performed using mini-batch gradient descent with a batch size of $B$, i.e., $B$ samples from the training dataset are processed together before updating the model weights. In all the results presented below, $B = 16$, as this provides a good compromise between memory usage and training stability. To increase robustness and generalization capabilities, a small Gaussian perturbation is applied to the normalized inputs parameters during training. This serves as a form of data augmentation, helping the model become more robust to small variations and reducing overfitting. During training, the model is evaluated every 10 epochs on the validation set using as validation error a point-wise Euclidean distance between the points of the predicted curve and the reference. The model achieving the best performance is saved.

\section{Results}\label{sec:results}
\subsection{Training results and model tuning}\label{sec:train_res}

\begin{figure}[h]
    \centering
    \includegraphics[width=0.45\linewidth]{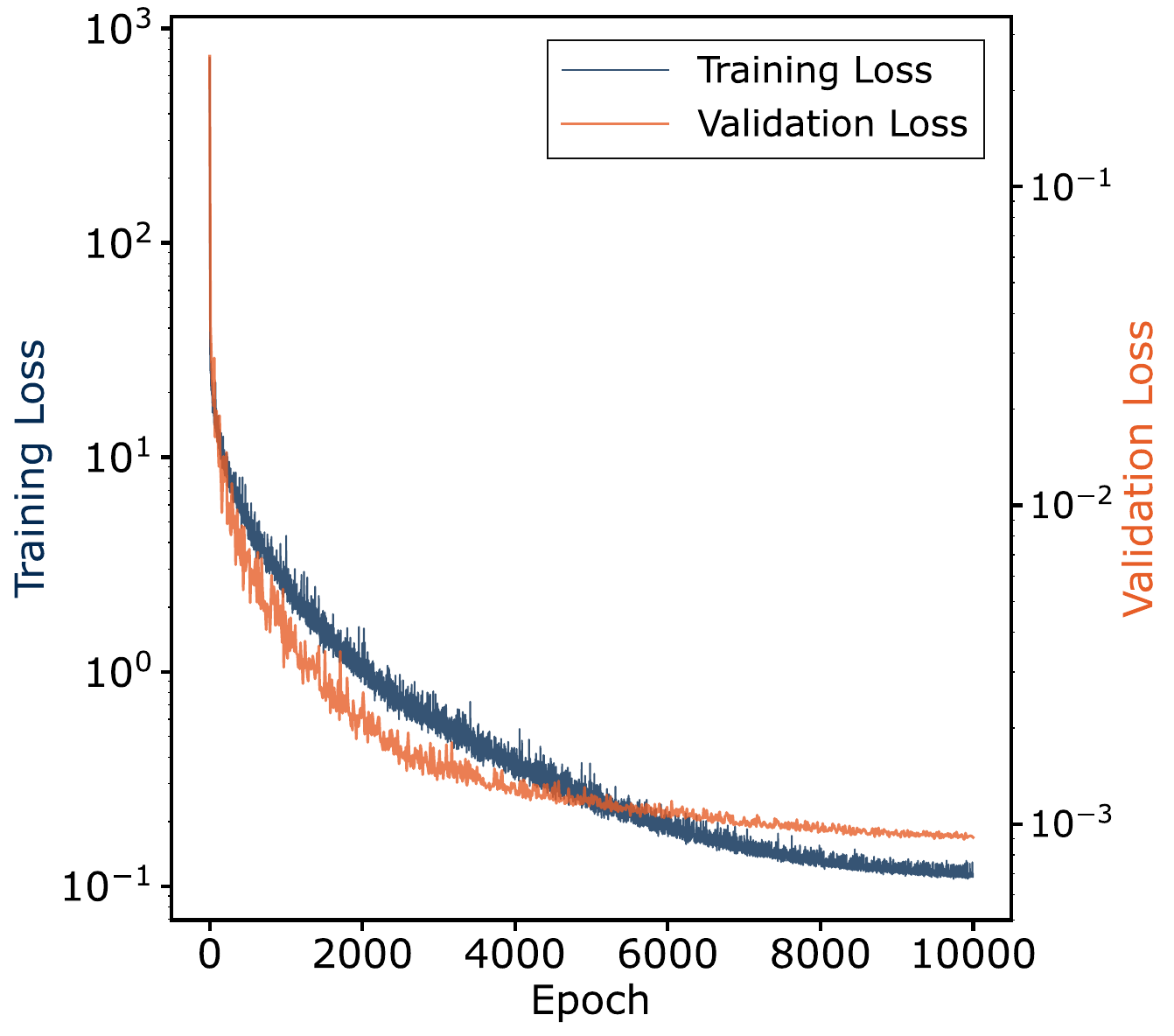}
    \caption{Training loss (blue) and validation error (orange) during training of the two-layer model. Both metrics are shown in logarithmic scale. The decline and convergence of the curves indicate stable learning and good generalization.}
    \label{fig:5}
\end{figure}

Two separate networks were trained to predict single-layer and two-layer cross-sections, respectively. Figure \ref{fig:5} illustrates the evolution of training loss and validation error for the two-layer model. Since different metrics were used, each curve is plotted on its own logarithmic scale. Despite this, both curves exhibit a consistent downward trend. In particular, the validation error remains stable and does not increase in the later stages, suggesting that the model generalizes well without evidence of overfitting. Although not shown, the learning curves for the single-layer model exhibited analogous behavior, with similar convergence trends and no indication of overfitting, confirming consistent generalization performance across both configurations.

\begin{figure}[h]
    \centering
    \includegraphics[width=\linewidth]{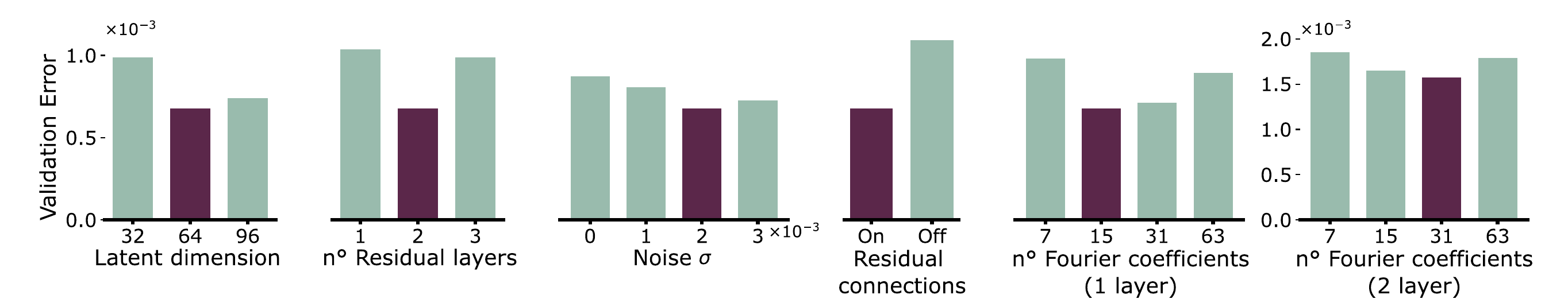}
    \caption{Validation error for different hyperparameter choices. Each subplot varies a single hyperparameter while keeping the others fixed. The purple columns are the final chosen sets of hyperparameters.}
    \label{fig:6}
\end{figure}

A hyperparameter sensitivity analysis was conducted to identify configurations that minimize validation error. The results, shown in Figure~\ref{fig:6}, report performance on the validation dataset across multiple training runs where each hyperparameter was varied individually while others were fixed. The final parameter combination is highlighted in purple.

Regarding "latent dimension" and the "number of residual layers" the analysis reveals the importance of carefully balancing model complexity. These components must be large enough to provide adequate representational power for capturing the complex, nonlinear mapping from input parameters to output curves. However, an excessive capacity increases the risk of overfitting, negatively impacting generalization.

As described in Section~\ref{subsec:network}, Gaussian noise is added to the input parameters during training as a form of data augmentation. This perturbation enhances the model’s robustness to small variations in the input space and improves its generalization performance. The magnitude of this perturbation must be carefully tuned, as large standard deviations ($\sigma$) can distort the data excessively. Furthermore, models with residual connections consistently achieved lower validation errors and enhanced training stability. These connections enhance representational power without compromising trainability, allowing the network to learn more complex transformations effectively.

Finally, the two bar plots on the right illustrate the effect of varying the number of Fourier coefficients used to parameterize the predicted curves. As discussed in Section~\ref{subsec:fourier}, selecting an appropriate number of Fourier coefficients is essential for accurately representing cross-section shapes. In the case of a single-layer geometry, where the shape is relatively simple, using 15 coefficients is sufficient to achieve good agreement with the ground truth. However, in the more complex two-layer case, which features a narrowing at the contact interface between layers, at least 31 coefficients are required to adequately capture geometric details.

\begin{figure}[h]
    \centering
    \subfloat[$\widetilde{N}=9$\label{fig:7a}]{\includegraphics[width=0.25\linewidth]{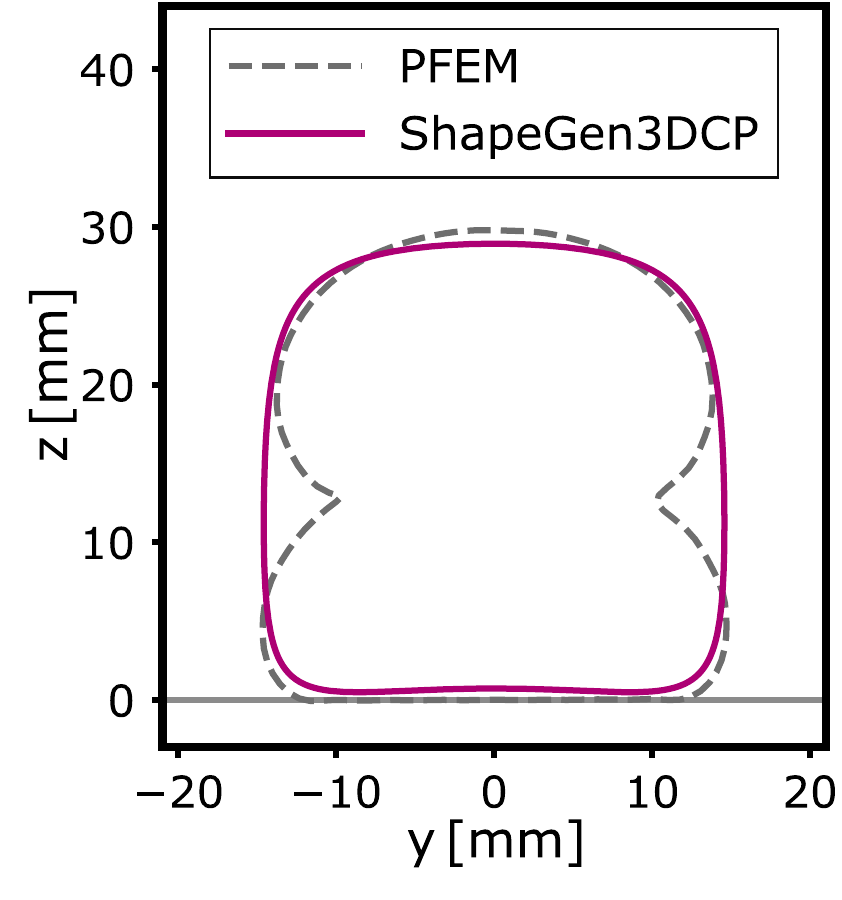}} 
    \subfloat[$\widetilde{N}=15$\label{fig:7b}]{\includegraphics[width=0.25\linewidth]{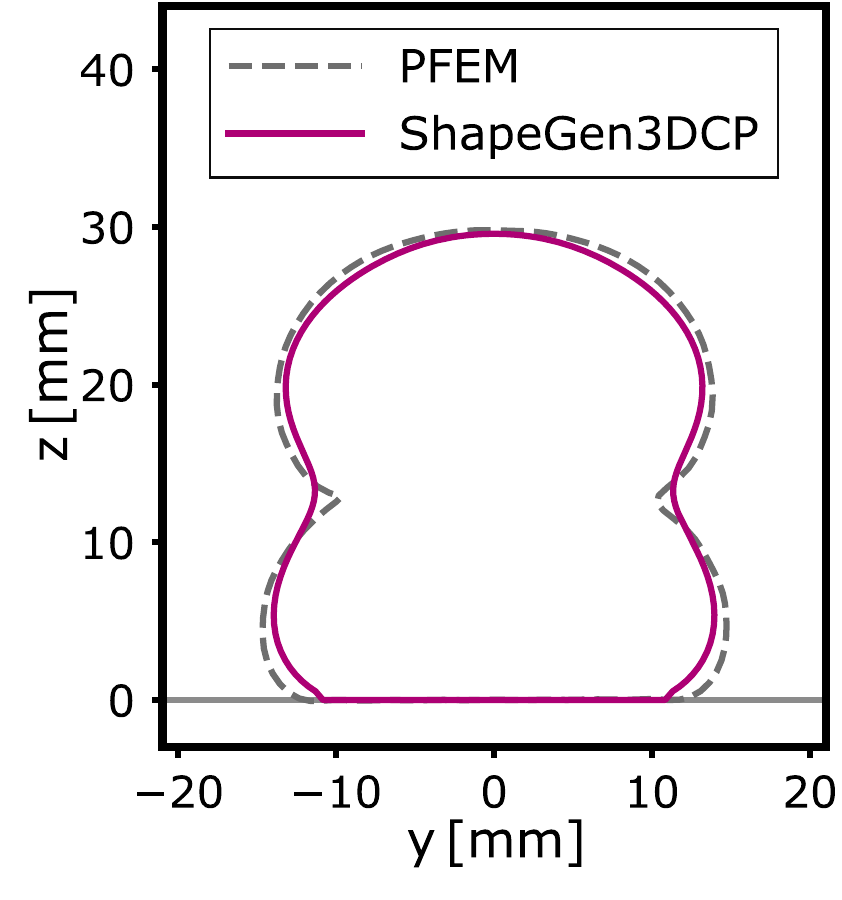}}
    \subfloat[$\widetilde{N}=31$\label{fig:7c}]{\includegraphics[width=0.25\linewidth]{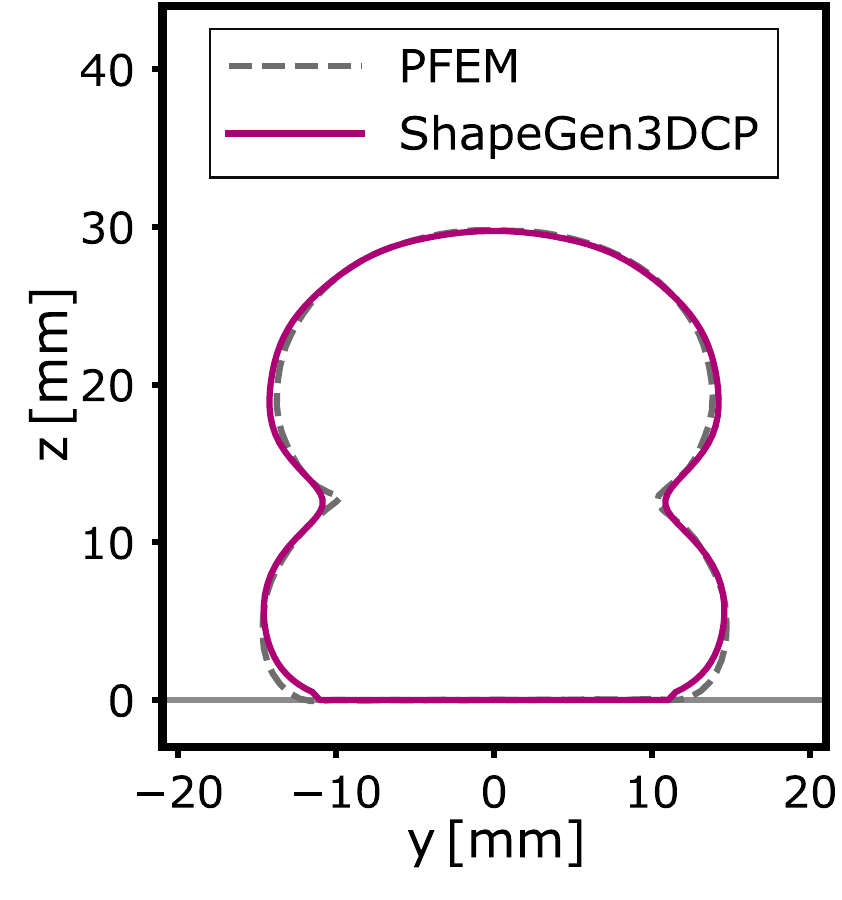}}
    \subfloat[$\widetilde{N}=63$\label{fig:7d}]{\includegraphics[width=0.25\linewidth]{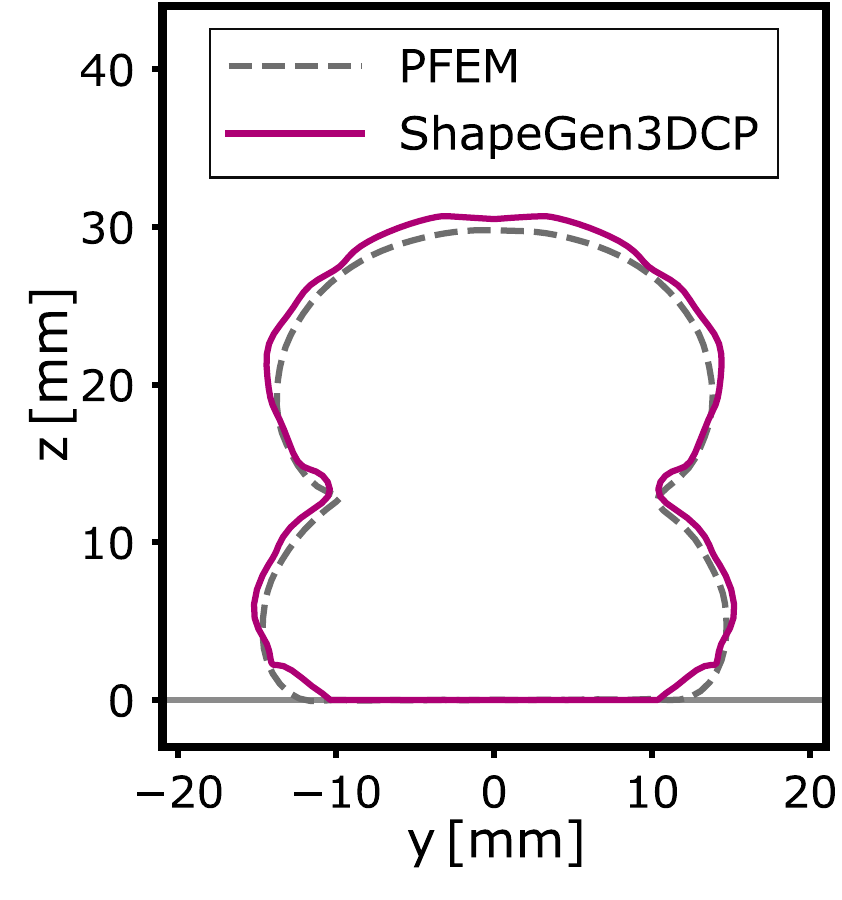}}
    \caption{Results obtained with models trained with varying numbers of Fourier coefficients.}
    \label{fig:7}
\end{figure}

Section~\ref{subsec:fourier} showed that directly fitting more Fourier coefficients drives convergence to the reference shape; however, a different pattern is observed when the coefficients are learned via the neural network. In this setting, using too many coefficients ($\widetilde{N}=63$) results in degraded performance, both for single and two-layer cases. 
The reason for this behavior is further illustrated in Figure~\ref{fig:7}, which shows predicted curves for different values of $\widetilde{N}$ in the two-layer case. With low-frequency representations, the model produces overly smooth shapes that fail to capture fine geometric details (Figure~\ref{fig:7a}-~\ref{fig:7b}). As $\widetilde{N}$ increases, the predictions improve, with $\widetilde{N}=31$ offering a good balance between fidelity and smoothness (Figure~\ref{fig:7c}). At $\widetilde{N}=63$, although the curve visually aligns with the reference, it exhibits high-frequency oscillations (Figure~\ref{fig:7d}). These results underscore the trade-off between representational capacity and regularity: while increasing $\widetilde{N}$ enhances expressiveness, it also raises the risk of learning unstable or noisy representations.

\newpage

\subsection{Performance evaluation and validation}
The effectiveness of the proposed method is first assessed by benchmarking its predictions against high-fidelity PFEM simulations not included in the training set. Specifically, the input parameters for ten selected numerical (N) cases are reported in Table \ref{tab:3}. These ten scenarios will be used to test the model’s accuracy in reproducing 3DCP cross-sections or relevant geometrical features across different parameter combinations to demonstrate the model's generalization capabilities.

\begin{table}[h]
\centering
\caption{Selected single and two layer cases for evaluating the predictive performance of \textit{ShapeGen3DCP} against unseen numerical data synthetically generated using the PFEM model proposed in \cite{rizzieri2024}.}
\label{tab:3}
\begin{tabular}{@{} c cccccccc @{}}
\toprule
& \textbf{ID} &
$\rho$ (kg/m$^3$) &
$\mu$ (Pa$\cdot$s) &
$\tau_0$ (Pa) &
$\phi_n$ (mm) &
$h_n$ (mm) &
$v_p$ (mm/s) &
$u_f$ (mm/s) \\
\midrule
\multirow{6}{*}{\textbf{1 Layer}}
& N1 & 2100   & 7.5   & 630    & 25     & 7.5    & 50    & 40.5 \\
& N2 & 2100   & 7.5   & 630    & 25     & 17.5   & 50    & 36.9 \\
& N3 & 2205   & 32.59 & 952.57 & 25.4   & 10     & 50    & 120   \\
& N4 & 2205   & 32.59 & 952.57 & 25.4   & 10     & 150   & 120   \\
& N5 & 2057.8 & 6.5   & 290.3  & 25.4   & 12.7   & 30    & 33.4  \\
& N6 & 2057.8 & 6.5   & 290.3  & 25.4 & 19.05  & 30    & 32.7  \\
\multirow{6}{*}{\textbf{2 Layers}}\\
& N7  & 2057.8 & 6.5   & 290.3  & 25.4   & 12.7  & 30   & 29.3  \\
& N8  & 2057.8 & 6.5   & 290.3  & 25.4   & 12.7  & 40   & 26.5  \\
& N9  & 2057.8 & 6.5   & 290.3  & 25.4   & 12.7  & 40   & 36.1  \\
& N10 & 2100   & 7.5   & 630    & 25     & 12.5  & 30   & 33.6  \\
\bottomrule
\end{tabular}
\end{table}

Succesively, to further validate \textit{ShapeGen3DCP}, a new set of predictions will be compared with experimental data reported by different research groups \cite{comminal2020,spangenberg2021,cheng2024,an2025}. The input parameters for the experimental validation cases are reported in Table \ref{tab:4}. This comparison will highlight the tool’s predictive accuracy and reliability, underlining its readiness for practical use by the 3DCP community.

\begin{table}[h]
\centering
\caption{Comparison between \textit{ShapeGen3DCP} predictions and numerically-generated data in terms of selected geometric features.}
\label{tab:4}
\begin{tabular}{@{} c cccccccc @{}}
\toprule
& \textbf{ID} &
$\rho$ (kg/m$^3$) &
$\mu$ (Pa$\cdot$s) &
$\tau_0$ (Pa) &
$\phi_n$ (mm) &
$h_n$ (mm) &
$v_p$ (mm/s) &
$u_f$ (mm/s) \\
\midrule
\multirow{6}{*}{\textbf{1 Layer}}
& E1, from \cite{comminal2020} & 2100   & 7.5   & 630    & 25     & 7.5    & 50    & 40.5 \\
& E2, from \cite{comminal2020} & 2100   & 7.5   & 630    & 25     & 12.5   & 30    & 33.6 \\
& E3, from \cite{cheng2024}    & 2205   & 32.59 & 952.57 & 25.4   & 15     & 50    & 120   \\
& E4, from \cite{cheng2024}    & 2205   & 32.59 & 952.57 & 25.4   & 20     & 75    & 120   \\
& E5, from \cite{an2025}       & 2057.8 & 6.5   & 290.3  & 25.4   & 12.7   & 40    & 36.3  \\
& E6, from \cite{an2025}       & 2057.8 & 6.5   & 290.3  & 25.4   & 19.05  & 30    & 32.7  \\
\multirow{6}{*}{\textbf{2 Layers}}\\
& E7, from \cite{an2025}       & 2057.8 & 6.5   & 290.3  & 25.4   & 12.7   & 30    & 35.1  \\
& E8, from \cite{an2025}       & 2057.8 & 6.5   & 290.3  & 25.4   & 12.7   & 40    & 26.5  \\
& E9, from \cite{an2025}       & 2057.8 & 6.5   & 290.3  & 25.4   & 12.7   & 40    & 36.1  \\
& E10, from \cite{spangenberg2021} & 2100 & 7.5 & 630    & 25     & 12.5   & 30    & 33.6  \\
\bottomrule
\end{tabular}
\end{table}

\subsubsection{Comparison with numerical data}
\begin{figure}[h]
			\centering
			\subfloat[Case N1]{\includegraphics[width=0.5\linewidth]{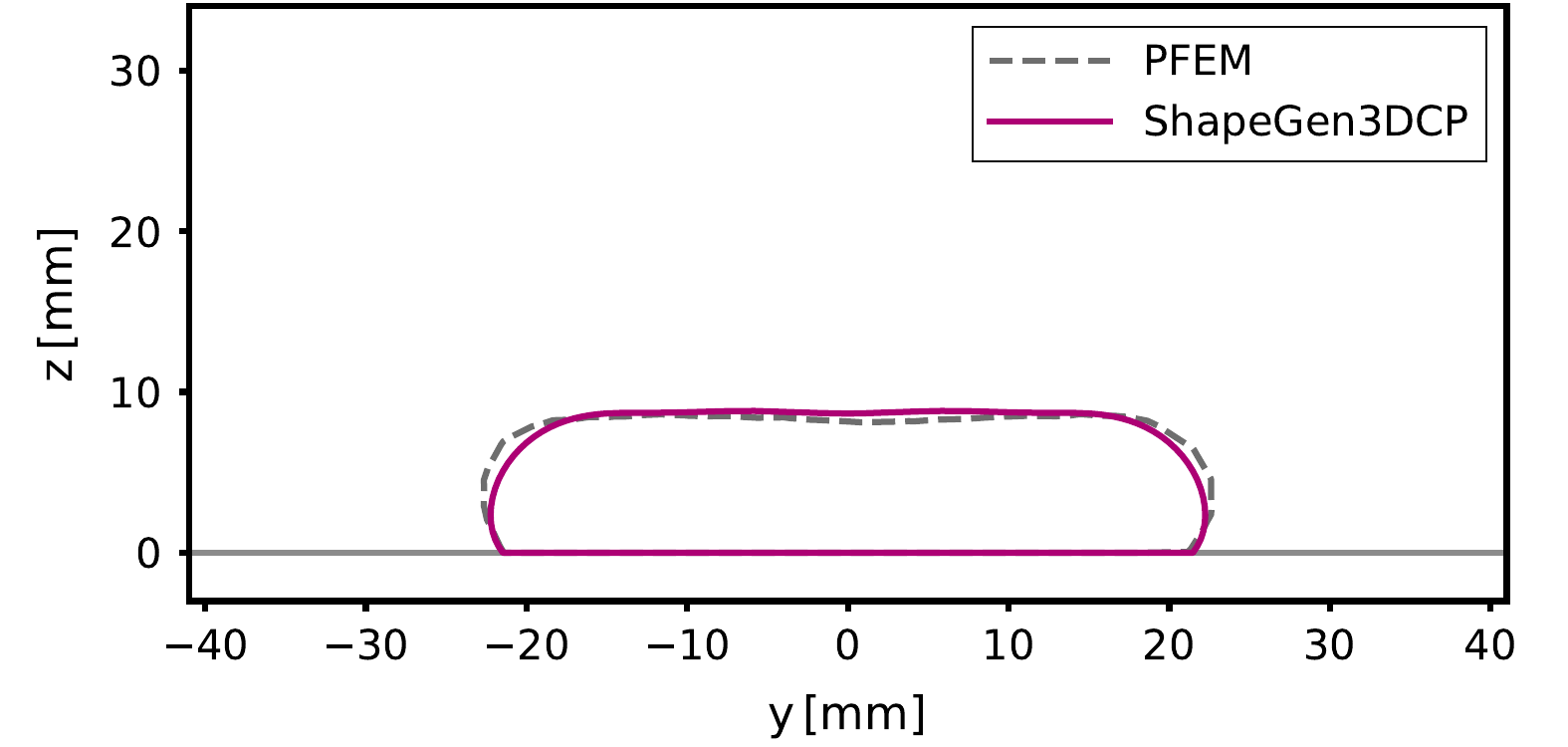}}
            \subfloat[Case N2]{\includegraphics[width=0.5\linewidth]{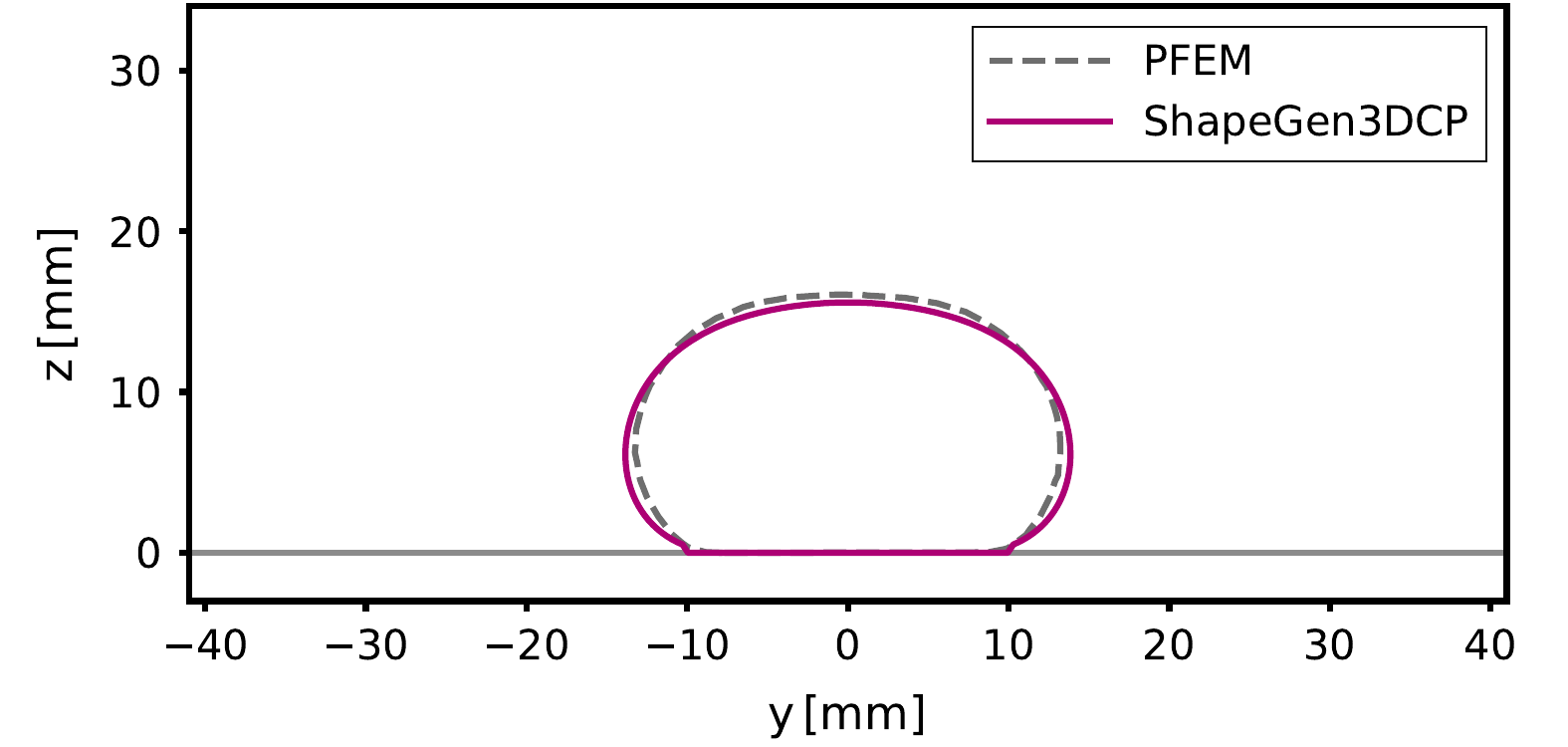}}\\
  		\subfloat[Case N3]{\includegraphics[width=0.5\linewidth]
            {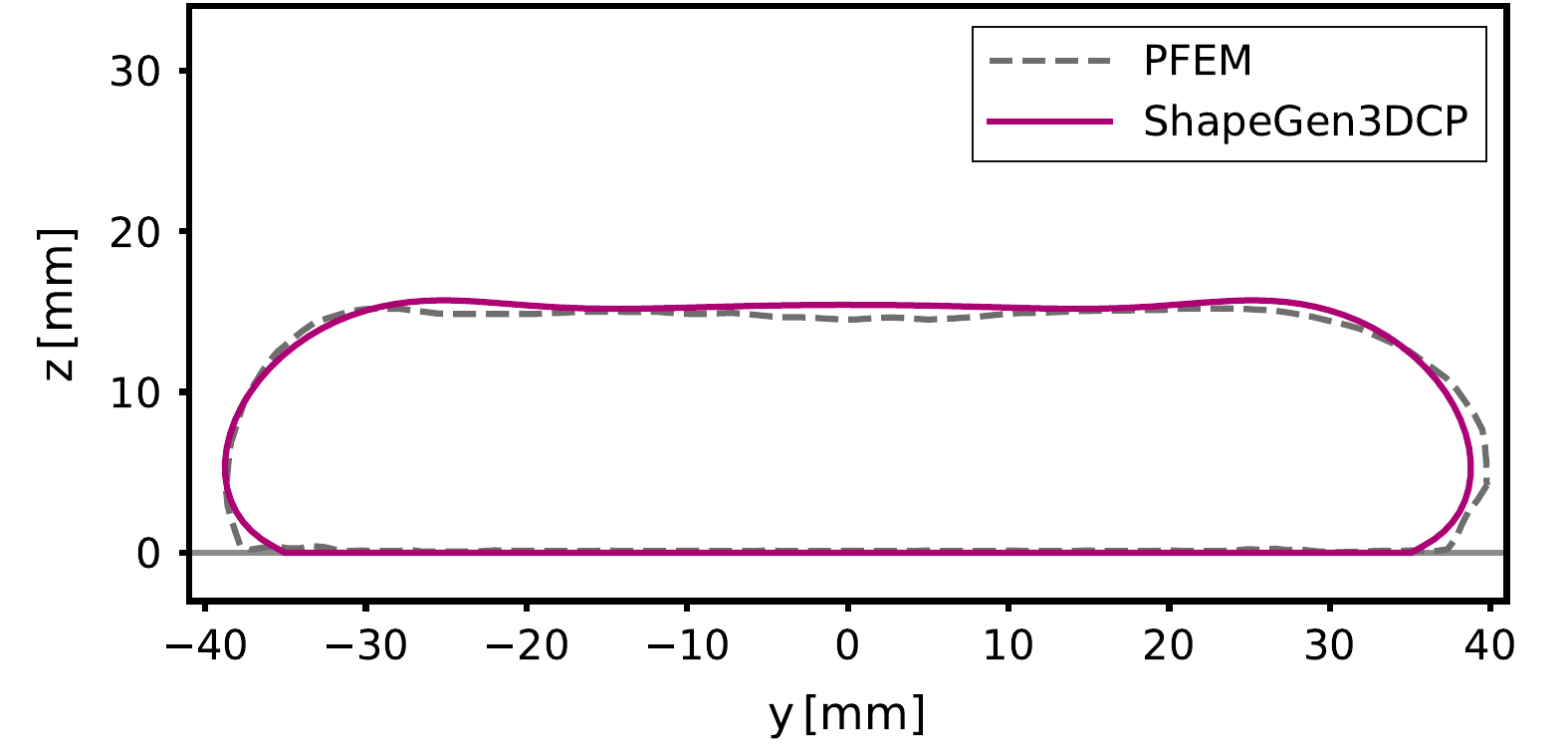}}
            \subfloat[Case N4]{\includegraphics[width=0.5\linewidth]{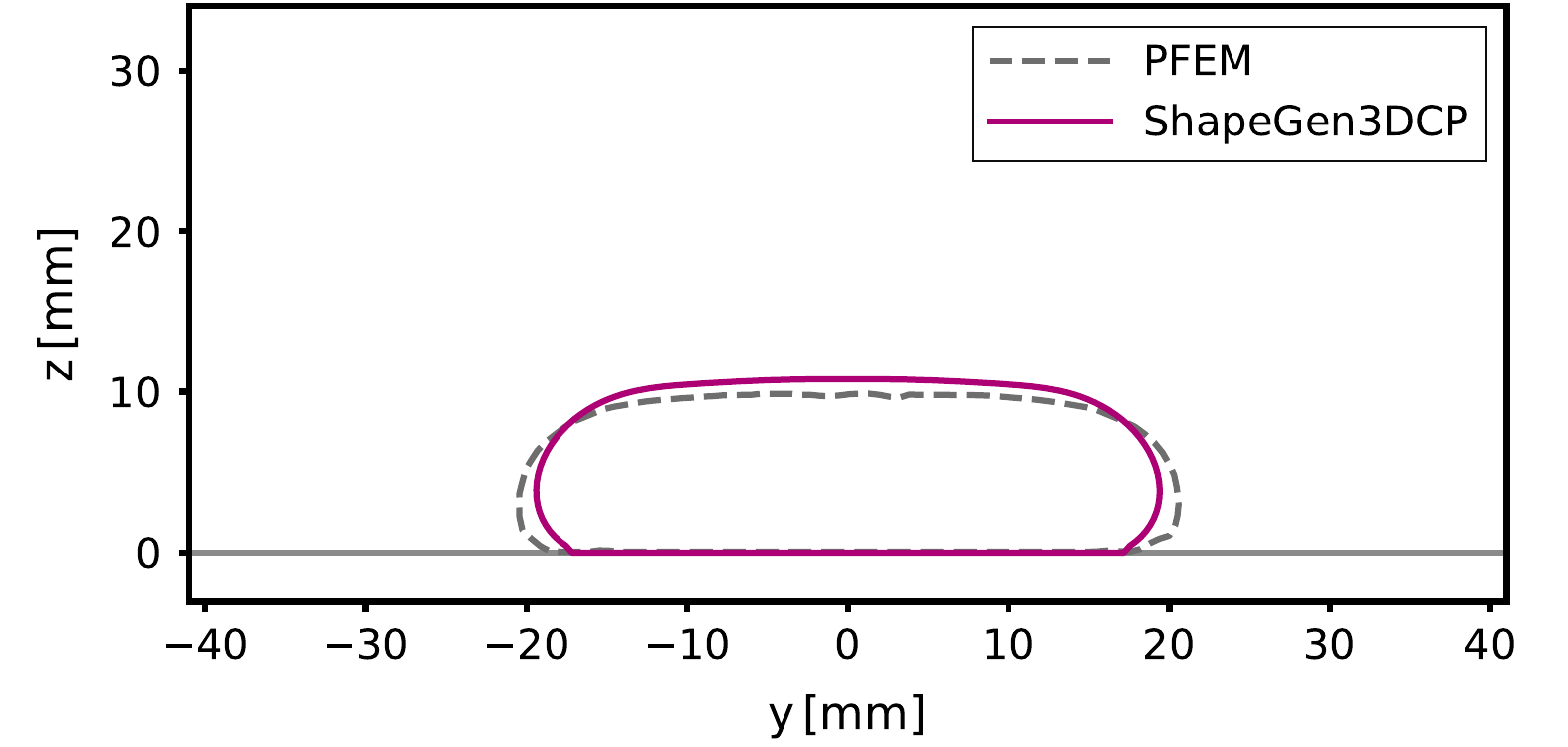}}\\
            \subfloat[Case N5]{\includegraphics[width=0.5\linewidth]{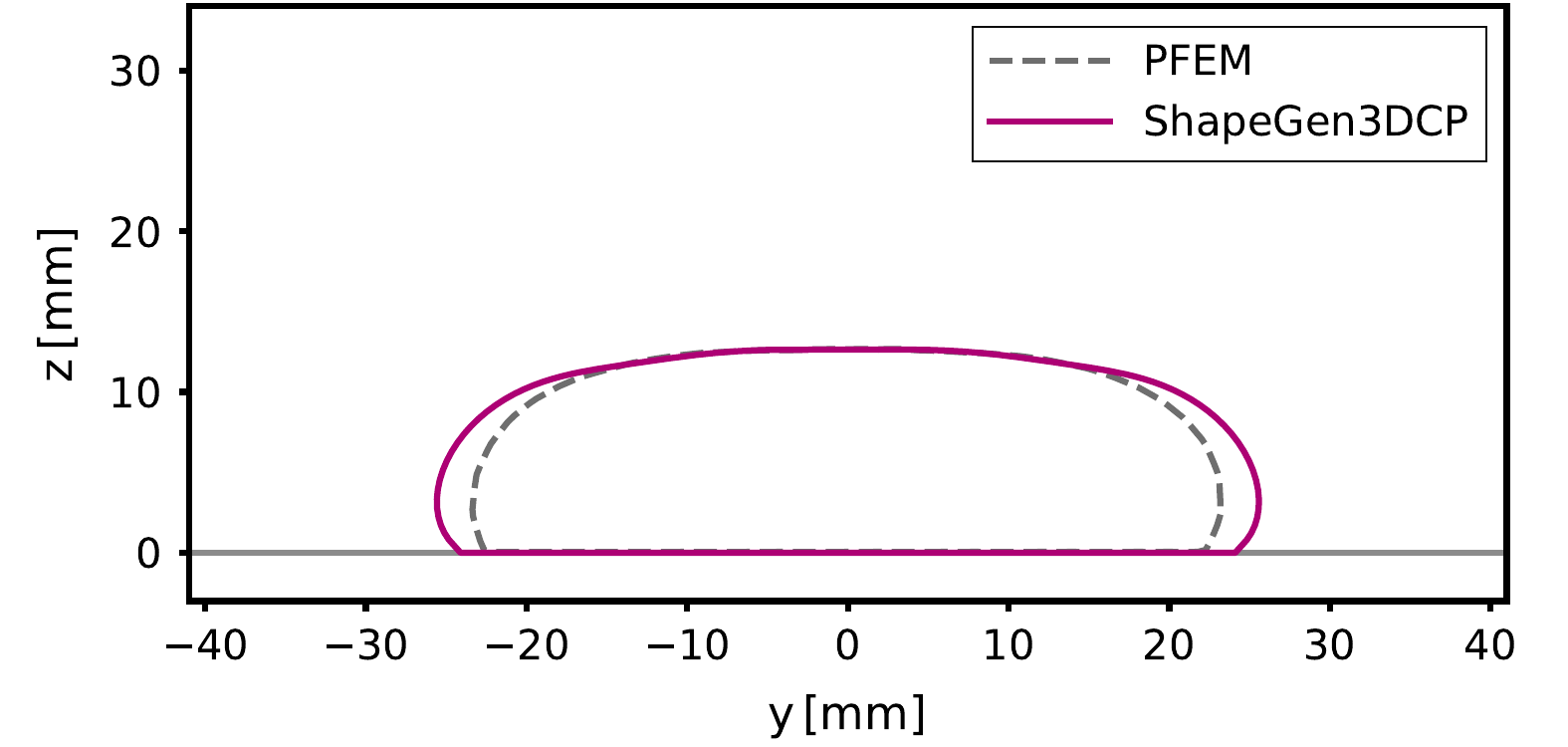}}
            \subfloat[Case N6]{\includegraphics[width=0.5\linewidth]{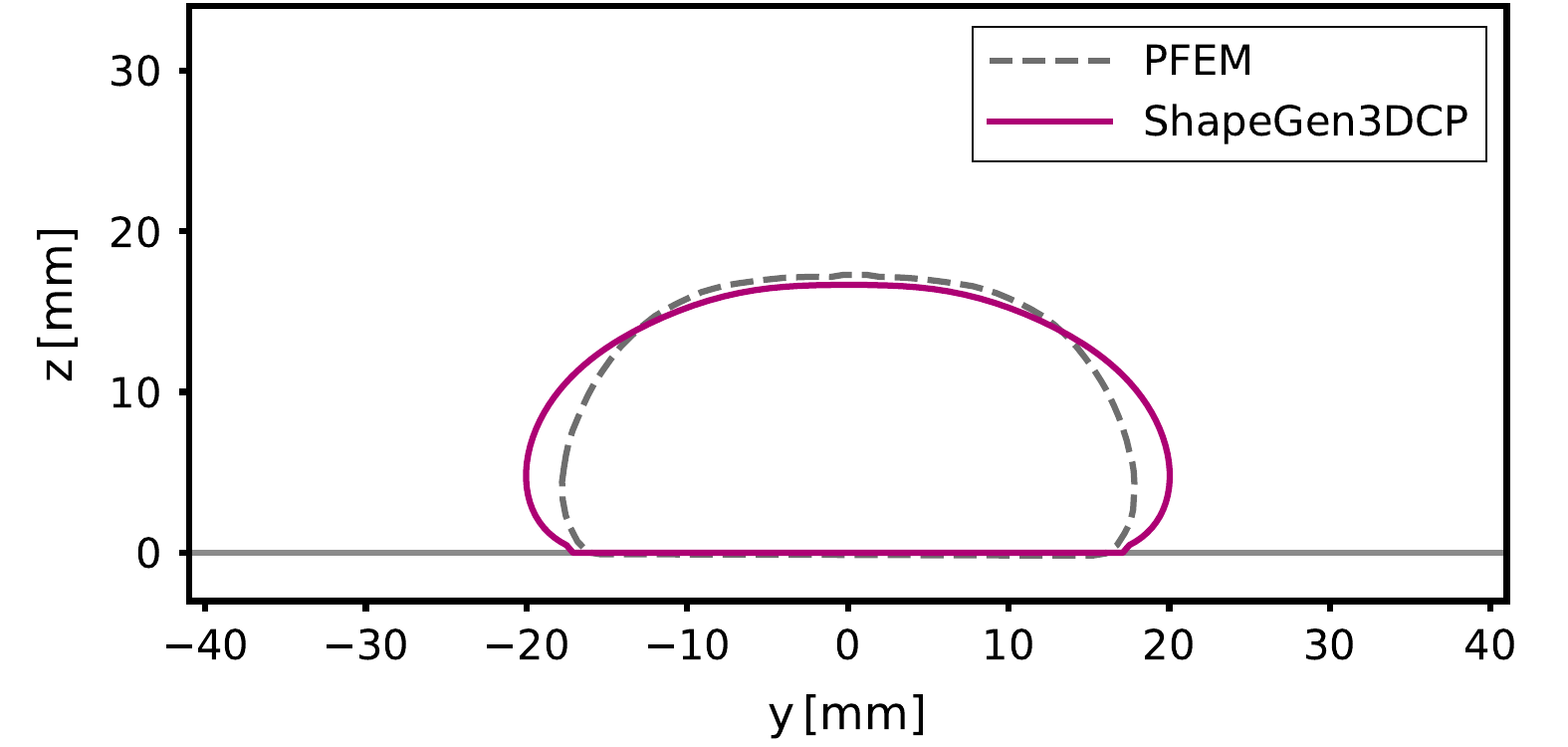}}
			\caption{Comparison of filament cross-sectional profiles predicted by \textit{ShapeGen3DCP} and those computed from numerical simulations for single-layer prints.}
			\label{fig:8}
		\end{figure}

\begin{figure}[h]
			\centering
			\subfloat[Case N7]{\includegraphics[width=0.5\linewidth]{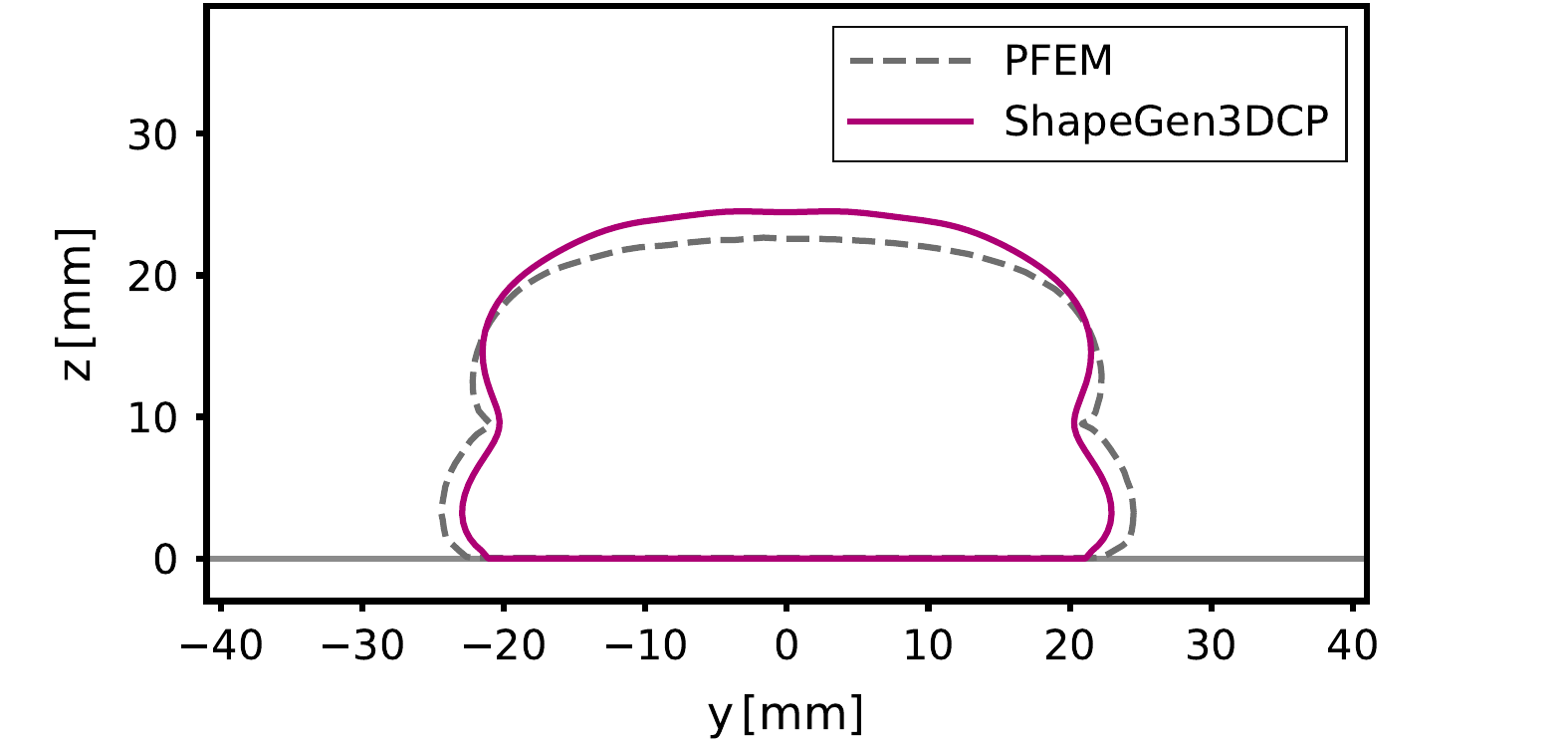}}
            \subfloat[Case N8]{\includegraphics[width=0.5\linewidth]{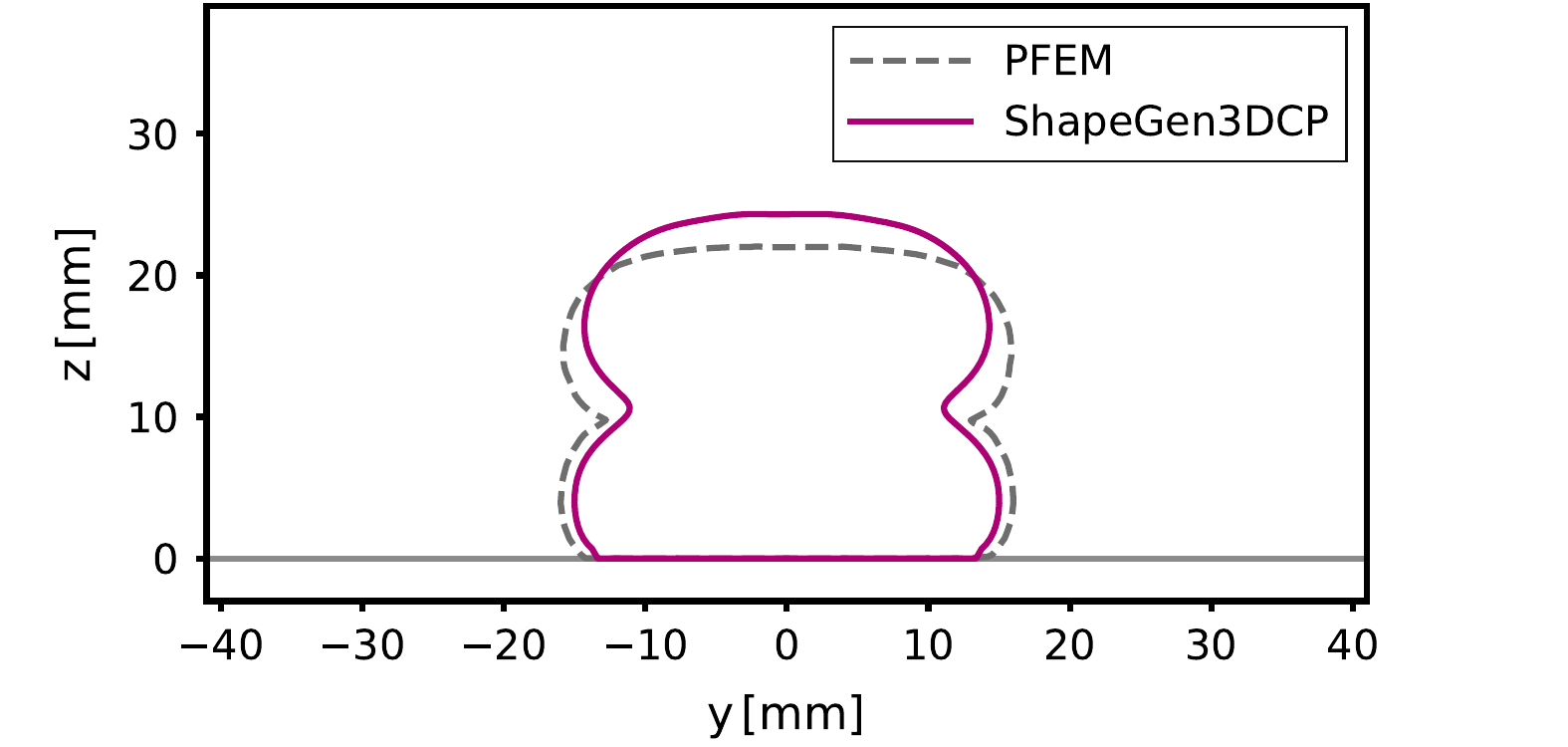}}\\
  		    \subfloat[Case N9]{\includegraphics[width=0.5\linewidth]{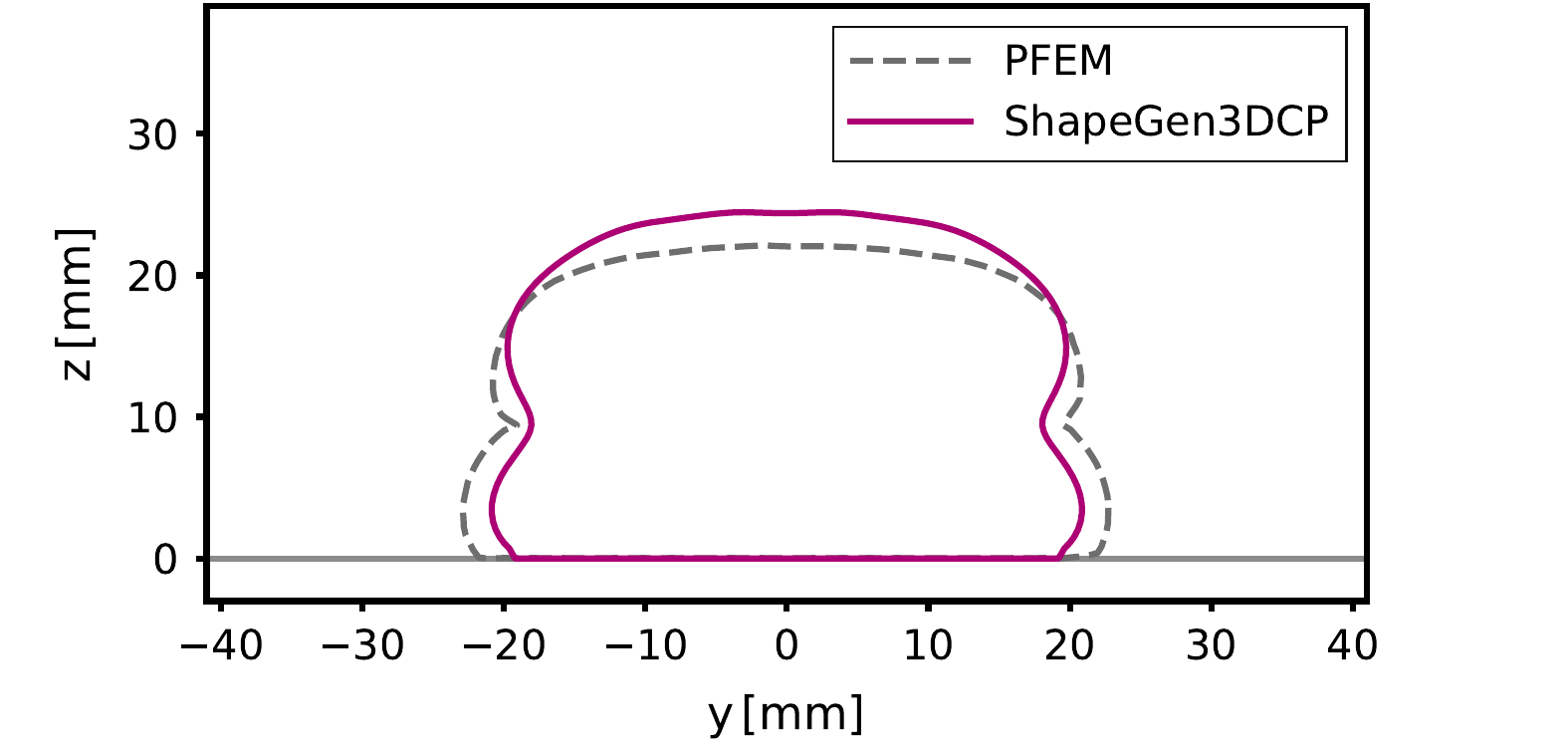}}
            \subfloat[Case N10]{\includegraphics[width=0.5\linewidth]{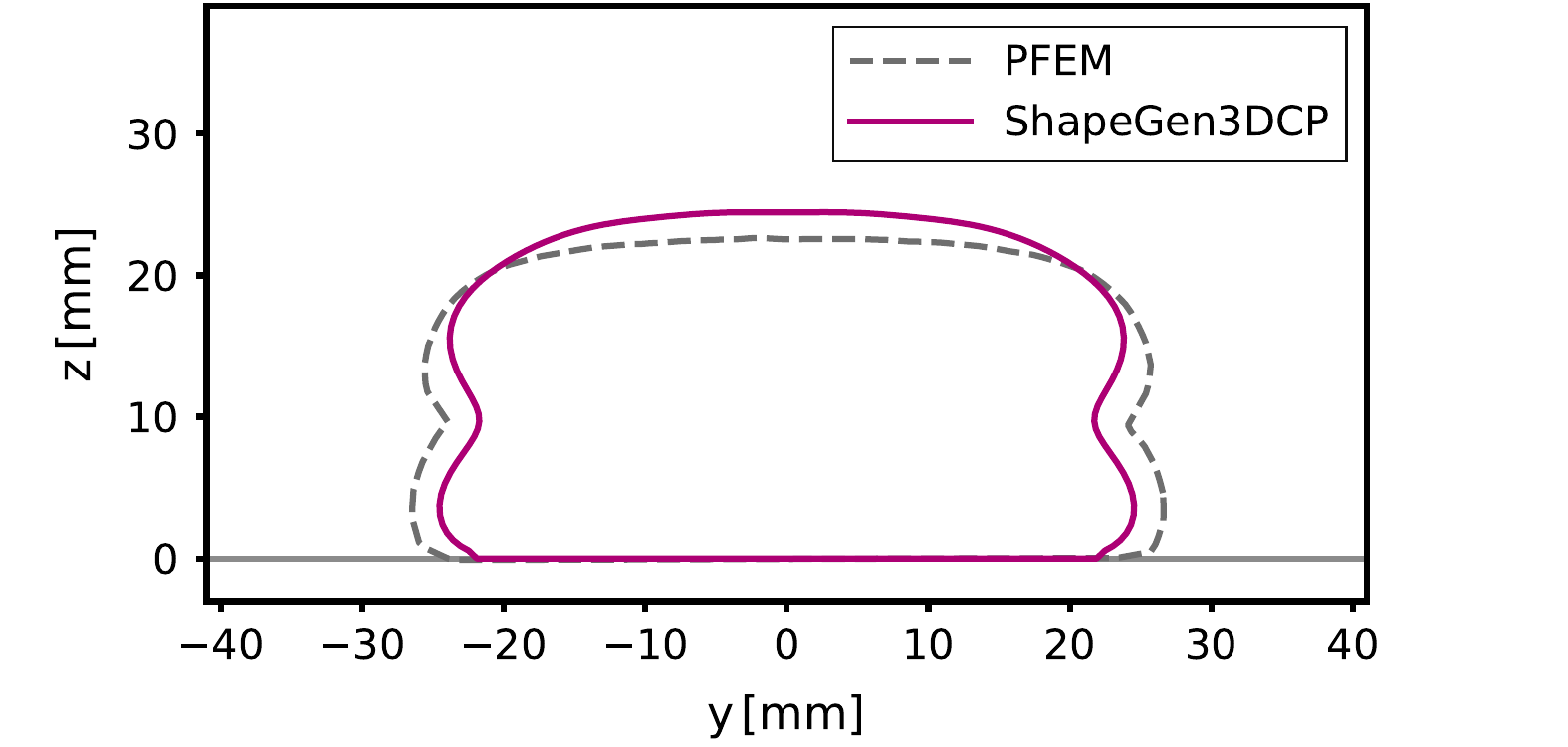}}
			\caption{Comparison of filament cross-sectional profiles predicted by \textit{ShapeGen3DCP} and those computed from numerical simulations for two-layers prints.}
			\label{fig:9}
		\end{figure}

Figure~\ref{fig:8} compares the cross-sections predicted by \textit{ShapeGen3DCP} with the ground-truth PFEM results for six single-layer cases (N1–N6), and Figure \ref{fig:9} shows similar comparisons for four two-layer cases (N7–N10). The input parameters for these cases are those previously reported in Table \ref{tab:3}.
  
Across all scenarios, the model demonstrates excellent agreement with the numerically simulated geometries. For single layers, the predictions accurately capture variations in width, height, and surface curvature resulting from different process conditions. In two-layer simulations, the model reliably reproduces the characteristic necking at the inter-layer interface, correctly capturing its vertical position and the extension of the contact zone. Minor discrepancies are observed primarily in regions of high curvature, but the overall shape fidelity is high.

To perform a deeper quantitative assessment, a set of relevant geometric features of interest for engineering applications has also been selected and used for comparison. With reference to the single layer or to the system composed of two overlapped layers, let $(x_i, y_i)$ denote the coordinates of a generic node $i$ on the profile. The selected features are the following: the maximum width $w = \max_{i,j} |x_i - x_j|$, the maximum height $h = \max_i y_i$, the cross-sectional area $A$ computed numerically using the shoelace formula, and, only for the two-layer case, the contact length $l_c$, obtained by detecting in post-processing the position of the pinch point, as the point where the approximate derivative ( evaluated via finite differences) changes sign. Other relevant features could also be extracted, such as the height of the first layer in the two-layer stack, which corresponds to the pinch-point height and reflects the vertical deformation of the first deposited layer. For clarity, the analysis is restricted to the set of features defined above.

The results comparing the features predicted by \textit{ShapeGen3DCP} with those extracted from the PFEM model are reported in Table~\ref{tab:5}, along with the corresponding percentage errors. Overall, \textit{ShapeGen3DCP} achieves good accuracy across all selected features, with relative errors generally in the range of $1 - 10 \%$. Among them, the cross-sectional areas are the most representative parameters for evaluating model performance, as they provide a global measure. For these, the errors are on average smaller than for the other features, which are instead more sensitive to local perturbations or numerical modeling artifacts. Nonetheless, regarding filament widths, heights, and the interlayer contact lengths, even in cases where the errors approach $10 \%$, the discrepancy corresponds to only 2–3 mm in practice, an uncertainty that is typically acceptable in additive manufacturing construction processes with cementitious materials.

Overall, the results confirm that the model generalizes well to unseen parameter combinations within the trained input domain, delivering accurate and reliable predictions of filament cross-sectional shapes as well as post-processed features.

\begin{table}[h]
\centering
\caption{Selected geometric features comparison between \textit{ShapeGen3DCP} predictions and numerically generated data.}
\label{tab:5}
\resizebox{\textwidth}{!}{%
\begin{tabular}{@{} c c | ccc | ccc | ccc | ccc @{}}
\toprule
& \textbf{ID} &
\makecell{$w_{SG}$ \\ (mm)} & \makecell{$w_{PFEM}$ \\ (mm)} & \makecell{$e_{w}$ \\ (\%)} &
\makecell{$h_{SG}$ \\ (mm)} & \makecell{$h_{PFEM}$ \\ (mm)} & \makecell{$e_{h}$ \\ (\%)} &
\makecell{$l_{c,SG}$ \\ (mm)} & \makecell{$l_{c,PFEM}$ \\ (mm)} & \makecell{$e_{l_c}$ \\ (\%)} &
\makecell{$A_{SG}$ \\ ($\text{mm}^2$)} & \makecell{$A_{PFEM}$ \\ ($\text{mm}^2$)} & \makecell{$e_{A}$ \\ (\%)} \\
\midrule
\multirow{6}{*}{\textbf{1 L}}
& N1 & 44.5 & 45.3 & 1.70 & 8.8 & 8.6 & 2.56 & -- & -- & -- & 366.1 & 365.1 & 0.26 \\
& N2 & 27.7 & 26.4 & 4.76 & 15.6 & 16.1 & 3.01 & -- & -- & -- & 365.9 & 363.9 & 0.54 \\
& N3 & 77.5 & 79.5 & 2.51 & 15.7 & 15.2 & 3.36 & -- & -- & -- & 1127.3 & 1102.7 & 2.23 \\
& N4 & 38.8 & 41.2 & 5.74 & 10.8 & 9.9 & 9.24 & -- & -- & -- & 378.2 & 364.6 & 3.72 \\
& N5 & 51.2 & 46.4 & 10.33 & 12.6 & 12.7 & 0.19 & -- & -- & -- & 567.7 & 518.5 & 9.49 \\
& N6 & 40.1 & 35.7 & 12.33 & 16.7 & 17.3 & 3.58 & -- & -- & -- & 561.2 & 530.5 & 5.79 \\
\multirow{6}{*}{\textbf{2 L}}\\
& N7 & 45.9 & 49.0 & 6.36 & 24.5 & 22.7 & 8.20 & 40.6 & 41.8 & 2.86 & 980.8 & 953.7 & 2.84 \\
& N8 & 30.0 & 32.0 & 6.21 & 24.3 & 22.0 & 10.40 & 22.2 & 26.0 & 14.6 & 636.6 & 647.6 & 1.70 \\
& N9 & 41.7 & 45.4 & 8.21 & 24.5 & 22.1 & 10.62 & 36.1 & 38.1 & 5.25 & 891.9 & 875.7 & 1.84 \\
& N10 & 49.1 & 53.2 & 7.79 & 24.5 & 22.6 & 7.99 & 43.5 & 48.4 & 10.1 & 1071.4 & 1092.8 & 1.97 \\
\bottomrule
\end{tabular}%
}
\end{table}

\newpage
\subsubsection{Comparison with experimental data}
\begin{figure}[h]
			\centering
			\subfloat[Case E1]{\includegraphics[width=0.5\linewidth]{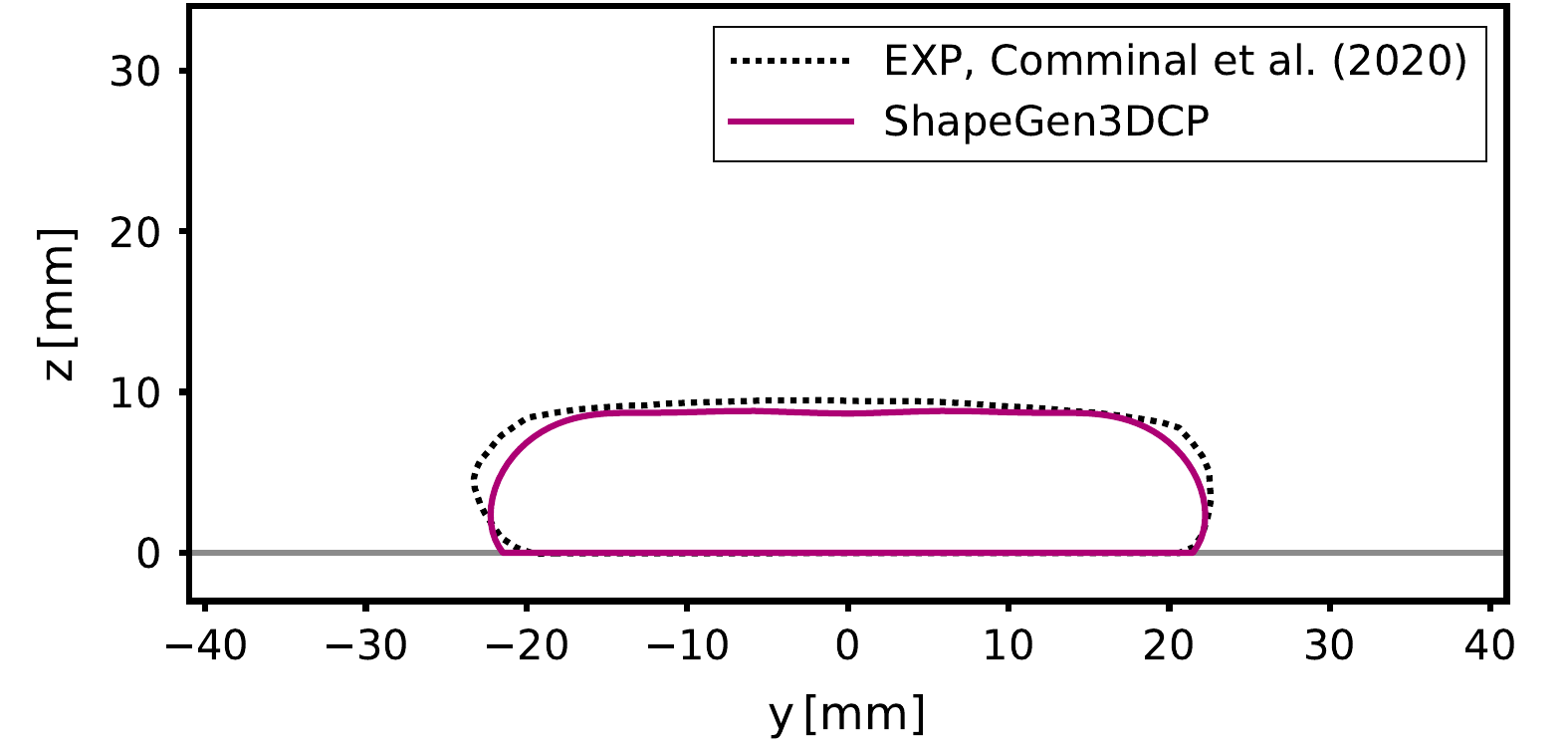}} 
            \subfloat[Case E2]{\includegraphics[width=0.5\linewidth]{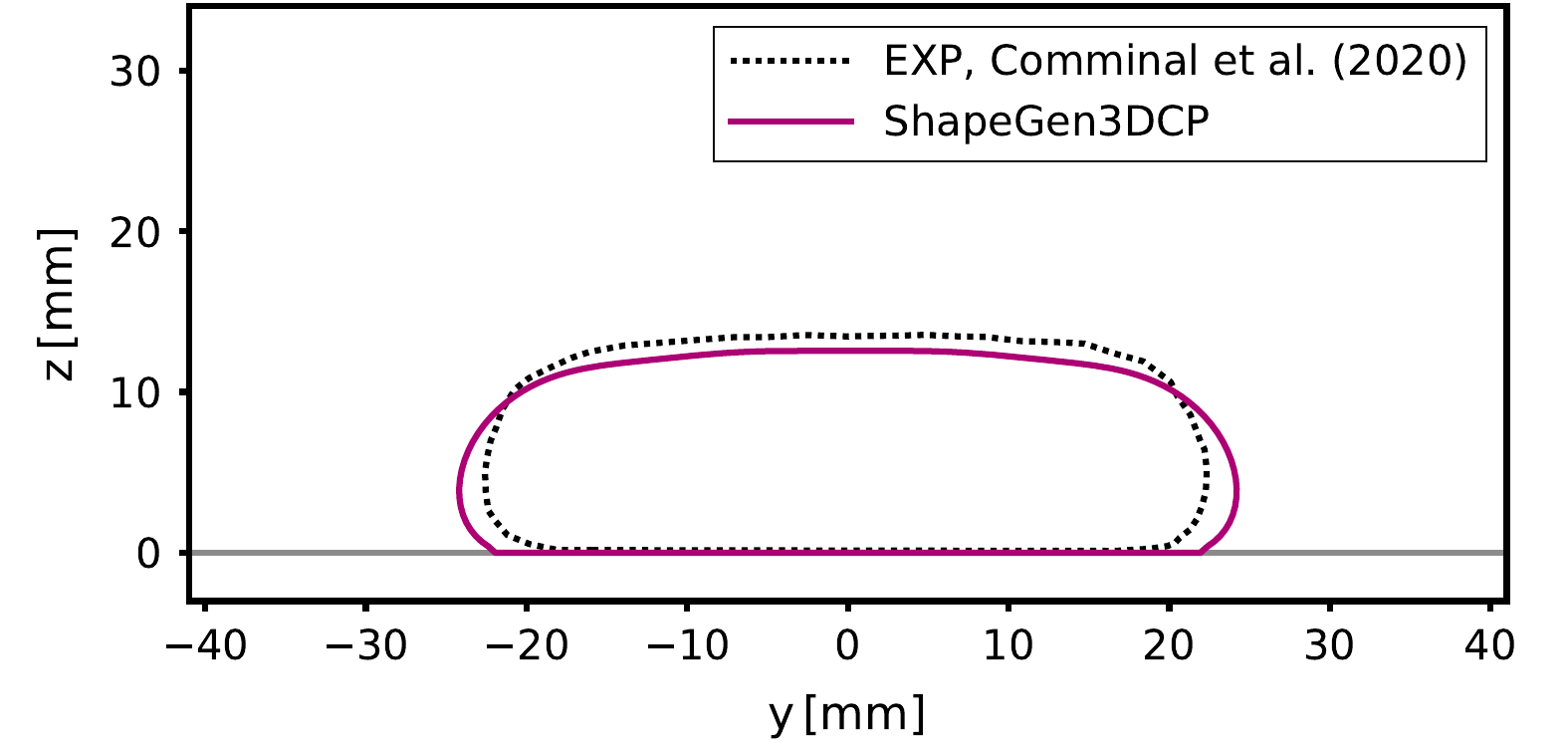}}\\
  		\subfloat[Case E3]{\includegraphics[width=0.5\linewidth]{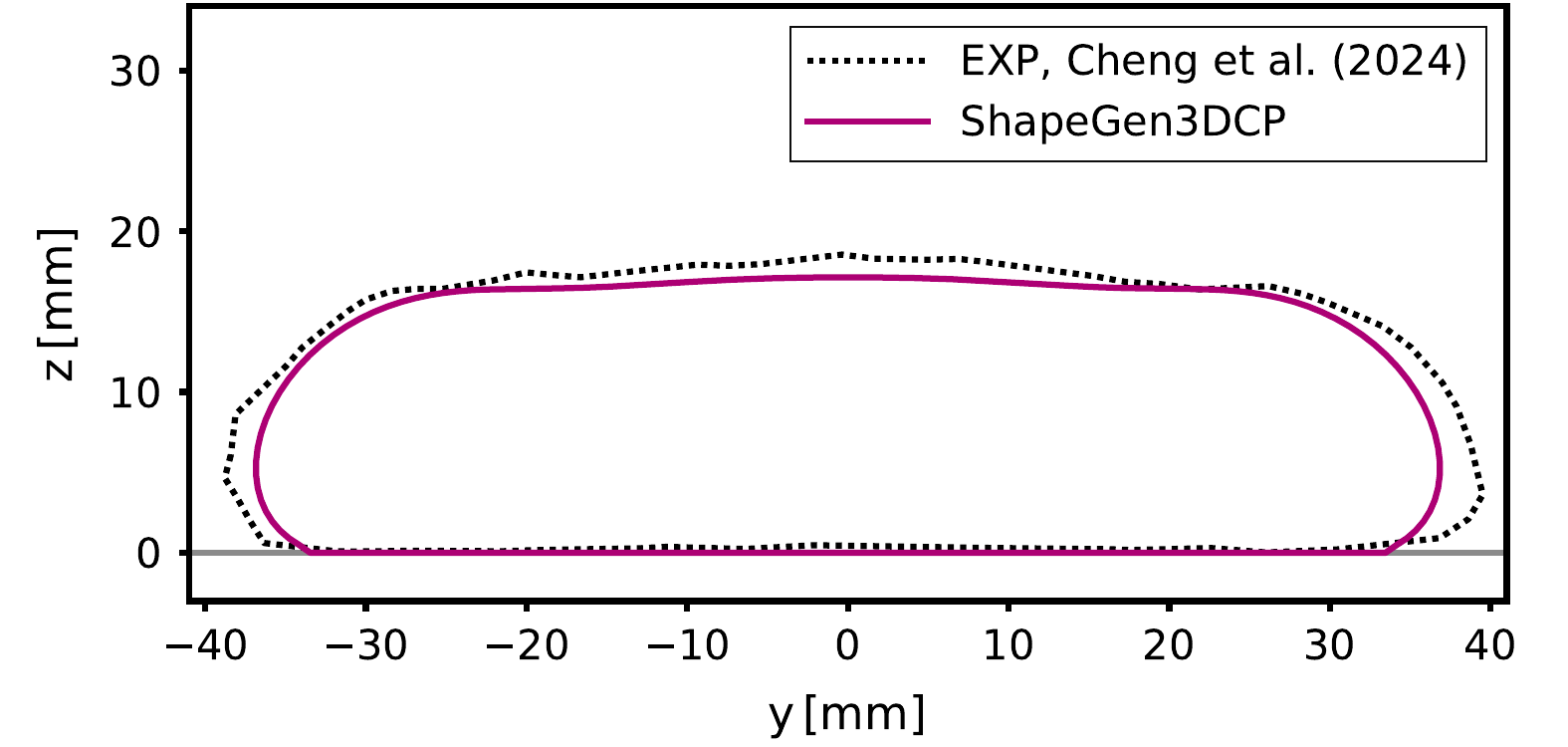}}
            \subfloat[Case E4]{\includegraphics[width=0.5\linewidth]{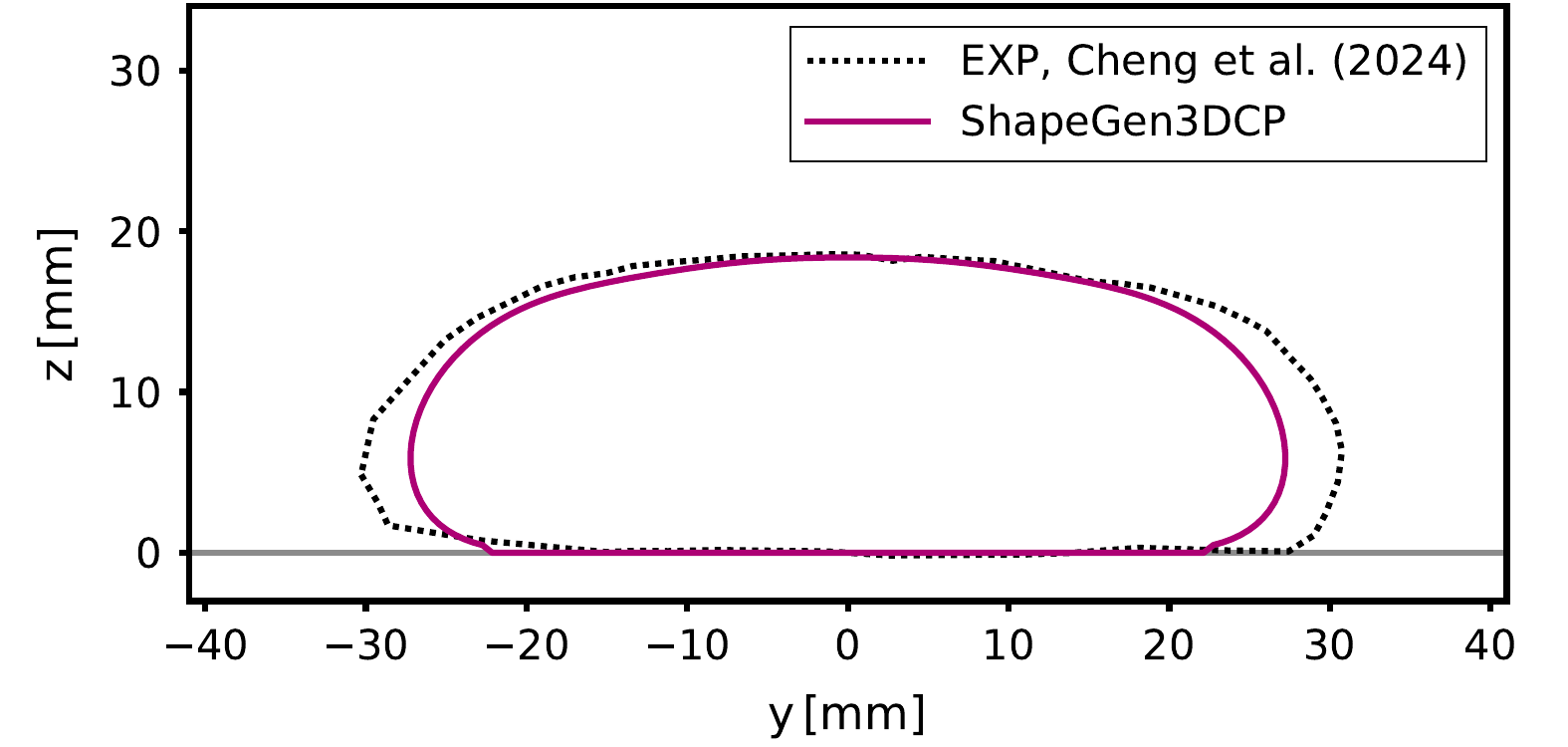}}\\
            \subfloat[Case E5]{\includegraphics[width=0.5\linewidth]{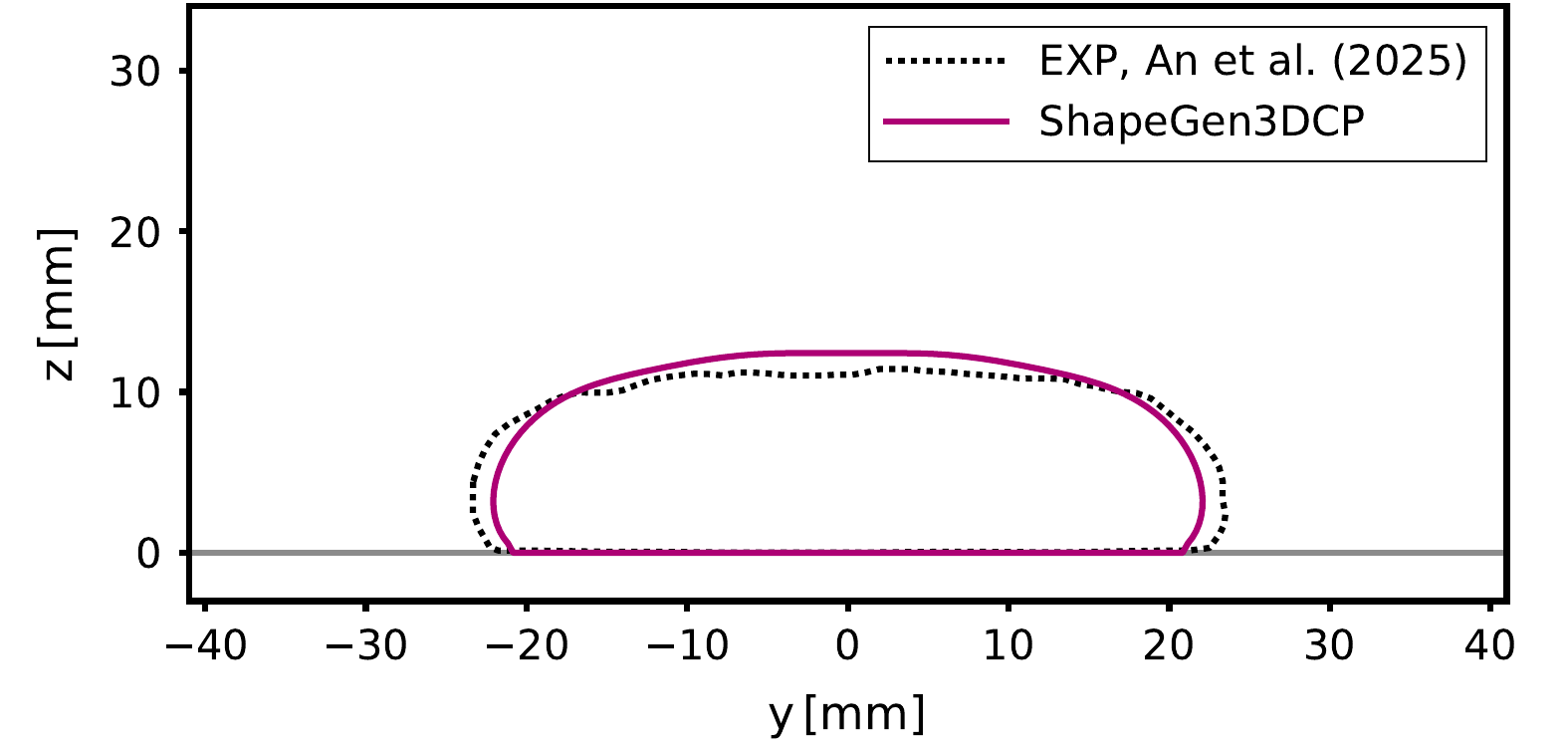}}
            \subfloat[Case E6]{\includegraphics[width=0.5\linewidth]{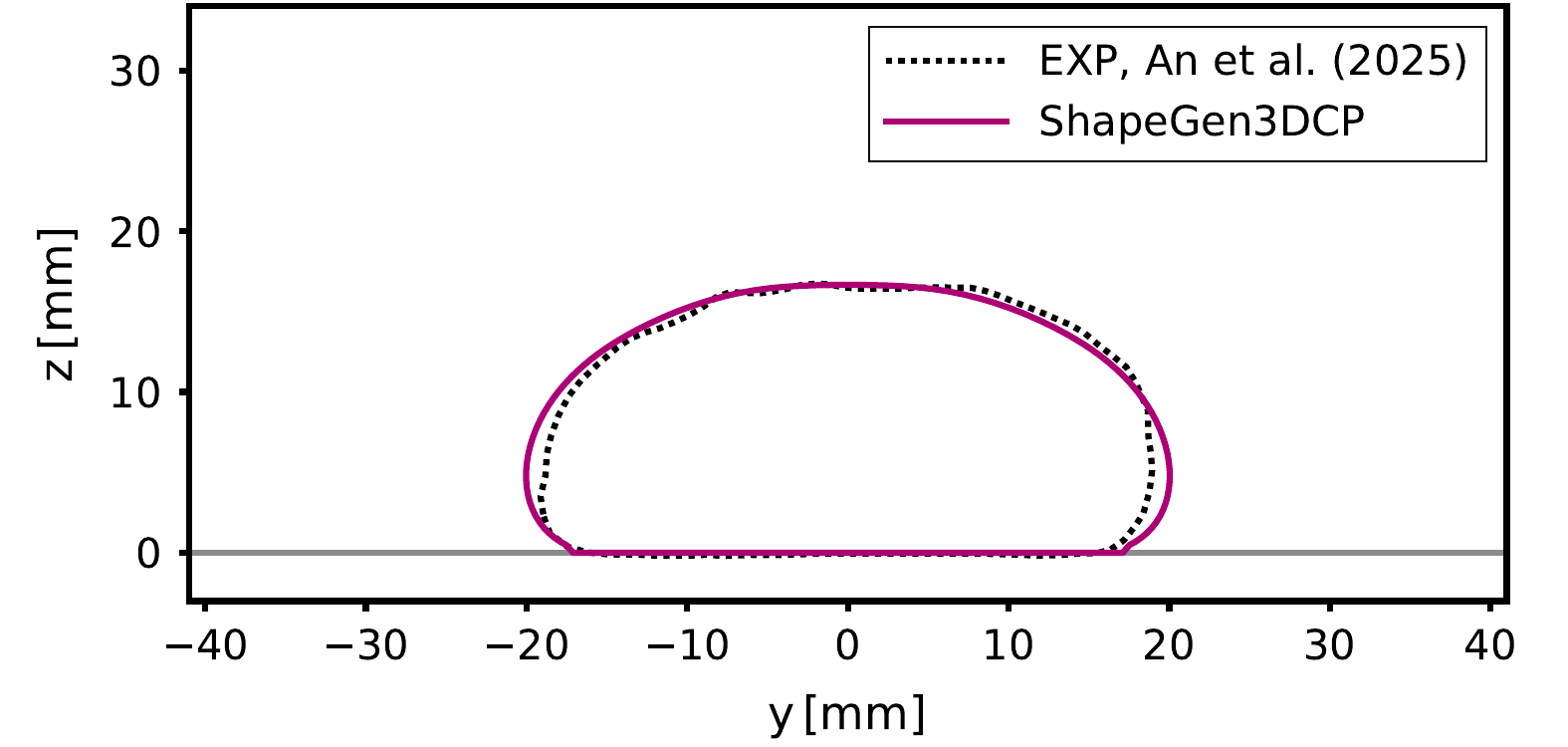}}
			\caption{Comparison of filament cross-sectional profiles predicted by \textit{ShapeGen3DCP} with experimental measurements from the literature for single-layer prints.}
			\label{fig:10}
		\end{figure}

\begin{figure}[h]
			\centering
			\subfloat[Case E7]{\includegraphics[width=0.5\linewidth]{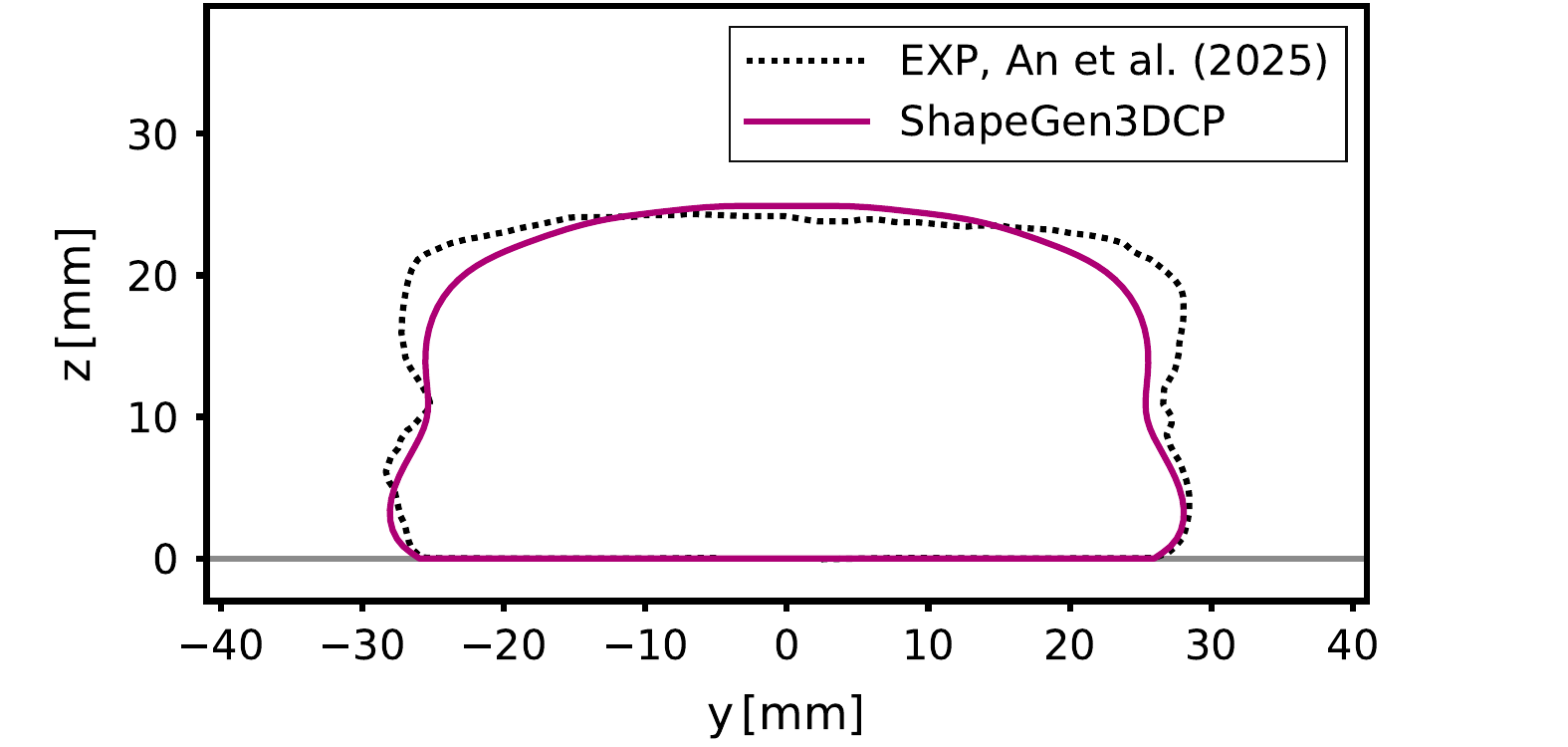}}
            \subfloat[Case E8]{\includegraphics[width=0.5\linewidth]{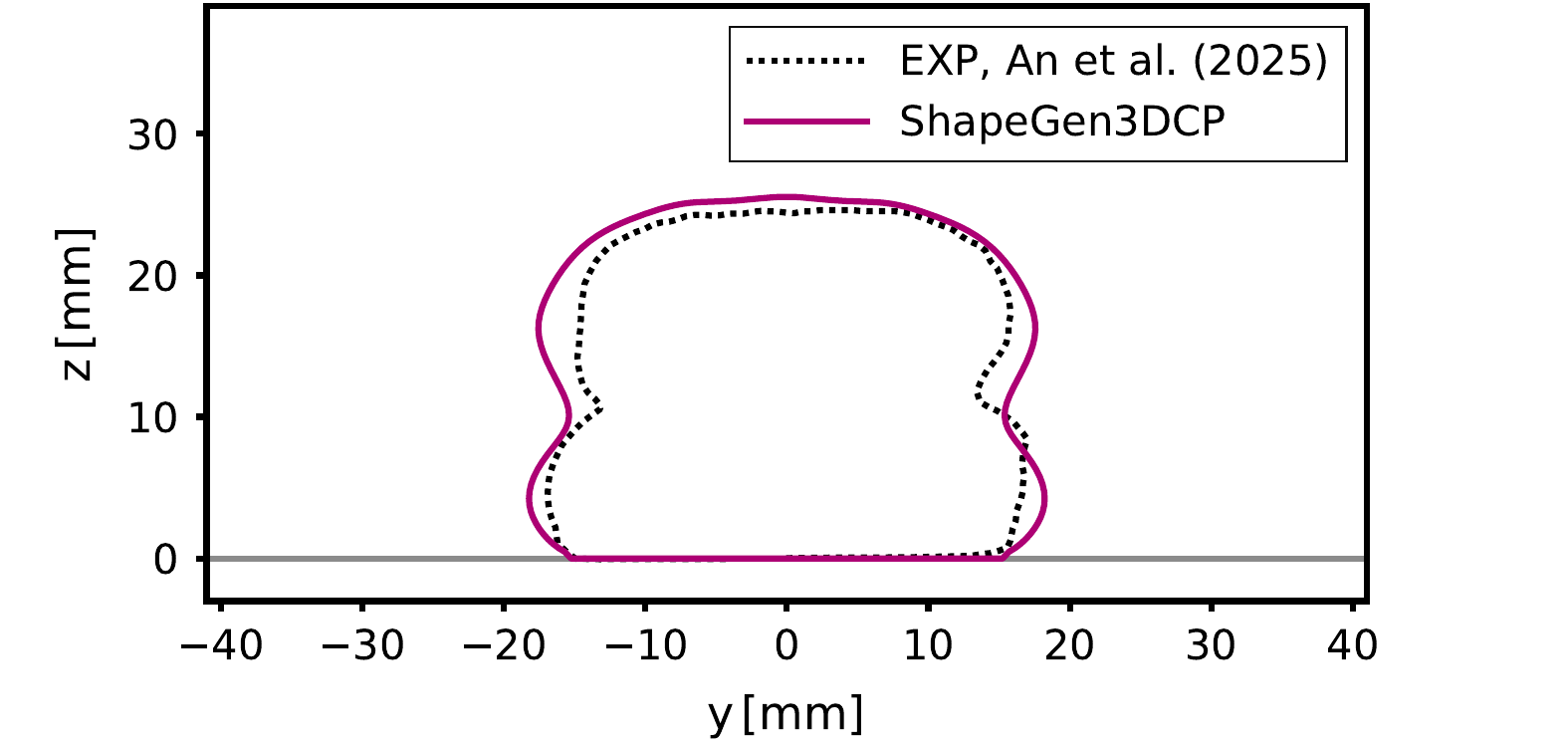}}\\
  		\subfloat[Case E9]{\includegraphics[width=0.5\linewidth]{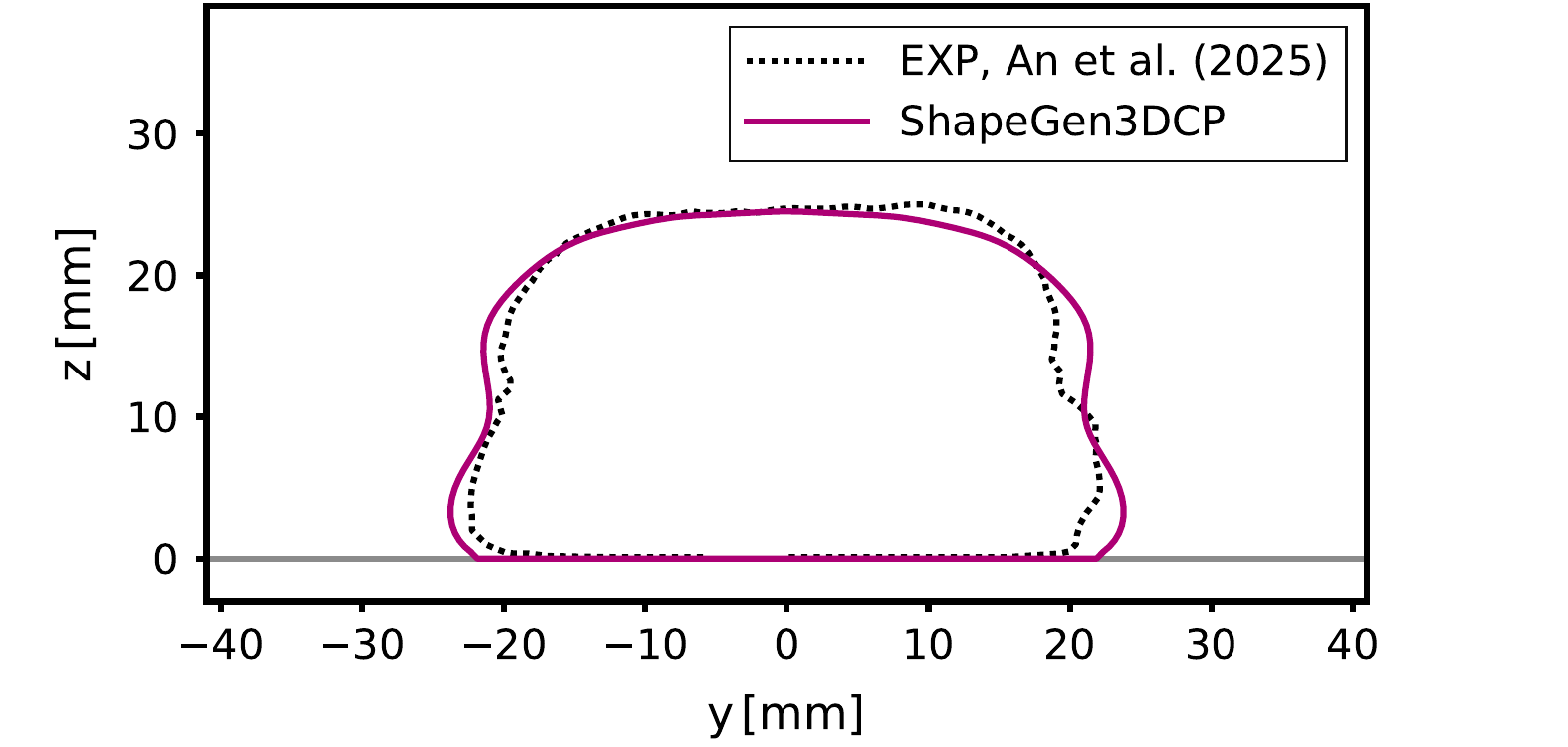}} 
            \subfloat[Case E10]{\includegraphics[width=0.5\linewidth]{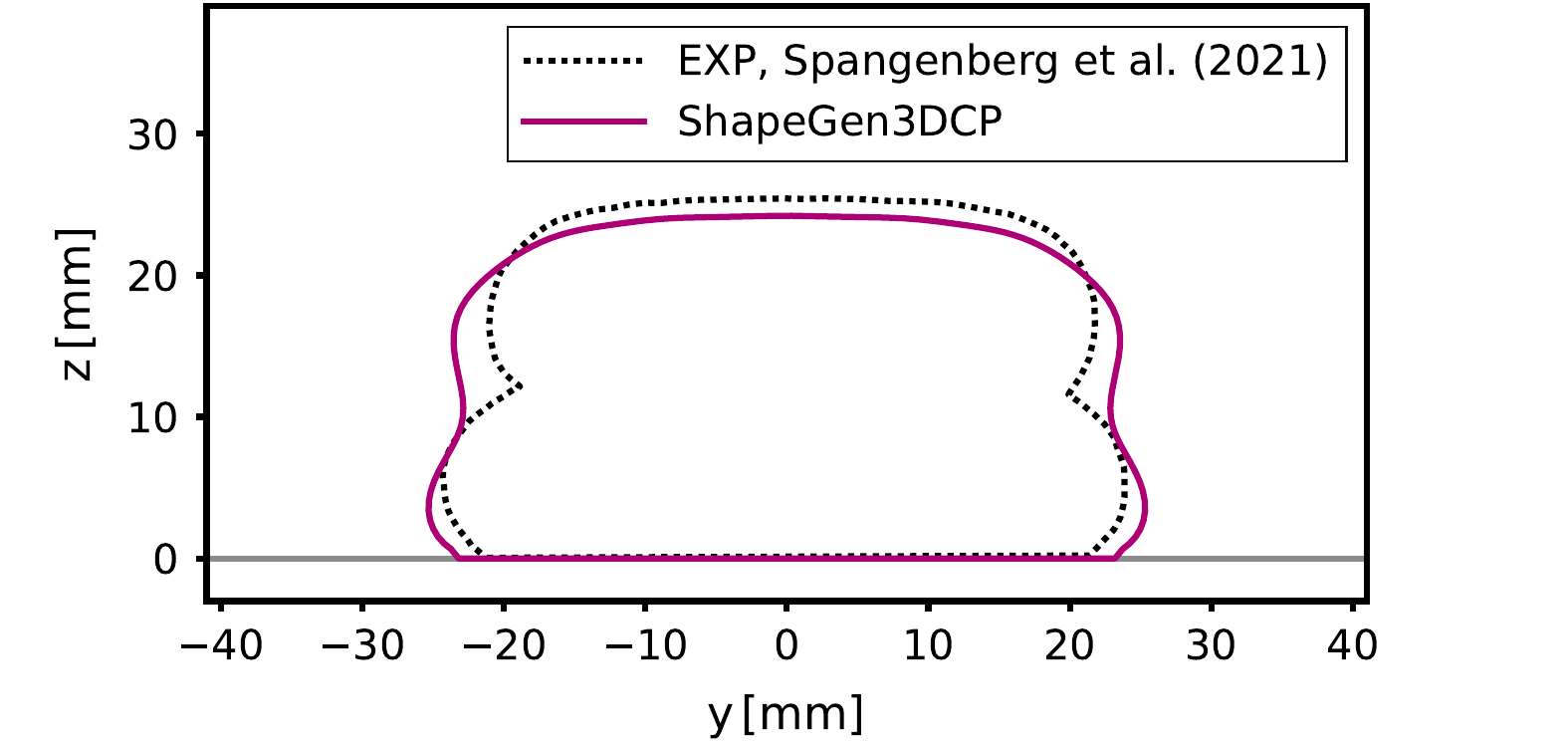}}
			\caption{Comparison of filament cross-sectional profiles predicted by \textit{ShapeGen3DCP} with experimental measurements from the literature for two-layer prints.}
			\label{fig:11}
		\end{figure}

To assess the real-world applicability of the tool, which it is worth recalling it was trained purely on numerically-generated data, a \textit{ShapeGen3DCP} performances were evaluated by predicting previously unseen experimental data provided by three independent research groups \cite{comminal2020,spangenberg2021,cheng2024,an2025}. These cases include the cross-sectional geometries of layers printed with different materials under varying printing conditions, as summarized in Table~\ref{tab:4}. In all cases, the cross-sections were obtained after or during hardening by cutting the specimens and reconstructing the contours using image recognition techniques. Specifically, Figures \ref{fig:10} and \ref{fig:11} compare the cross-sections predicted by \textit{ShapeGen3DCP} with the experimentally measured ones, for single-layer and two-layer scenarios, respectively.

Despite the inherent variability in material behaviour, printing settings, and experimental conditions, \textit{ShapeGen3DCP} consistently delivers accurate predictions. For instance, in cases E1 and E2 \cite{comminal2020}, the model correctly reproduces the characteristic filament flattening caused by layer pressing. In cases E4 and E6, characterized by rounder filaments due to free-flow deposition, it accurately captures both the curvature and the overall size. The model also demonstrates robustness to scale variations, successfully predicting wider filaments (e.g., E3 with $w \simeq 80$ mm) as well as smaller ones (e.g., E6 with $w \simeq 40$ mm).

For two-layer prints (E7–E10), the predicted profiles closely match experimental observations, including the interlayer contact zone and overall morphology, showcasing a wider bottom layer and a less spread upper layer. Notably, for all single- and two-layer cases, the contact length with the base plane, representative of gravity-induced spreading on the substrate, is also very well captured.

Also in this case, the comparison has performed in terms of relevant features, which were obtained by post-processing the predicted geometry, in the specific, the layer's width, height, area and for the two-layers case, the interlayer contact length. Results are reported with the relative percentage errors in Table \ref{tab:6}.

\begin{table}[h]
\centering
\caption{Selected geometric features comparison between \textit{ShapeGen3DCP} predictions and experimental data from the literature.}
\label{tab:6}
\resizebox{\textwidth}{!}{%
\begin{tabular}{@{} c c | ccc | ccc | ccc | ccc @{}}
\toprule
& \textbf{ID} &
\makecell{$w_{SG}$ \\ (mm)} & \makecell{$w_{EXP}$ \\ (mm)} & \makecell{$e_{w}$ \\ (\%)} &
\makecell{$h_{SG}$ \\ (mm)} & \makecell{$h_{EXP}$ \\ (mm)} & \makecell{$e_{h}$ \\ (\%)} &
\makecell{$l_{c,SG}$ \\ (mm)} & \makecell{$l_{c,EXP}$ \\ (mm)} & \makecell{$e_{l_c}$ \\ (\%)} &
\makecell{$A_{SG}$ \\ ($\text{mm}^2$)} & \makecell{$A_{EXP}$ \\ ($\text{mm}^2$)} & \makecell{$e_{A}$ \\ (\%)} \\
\midrule
\multirow{6}{*}{\textbf{1 L}}
& E1 & 44.48 & 45.18 & 1.56 & 8.83 & 9.50 & 7.06 & -- & -- & -- & 366.10 & 399.76 & 8.42 \\
& E2 & 48.39 & 44.64 & 8.39 & 12.56 & 13.55 & 7.32 & -- & -- & -- & 544.63 & 547.83 & 0.58 \\
& E3 & 73.67 & 79.02 & 6.77 & 17.14 & 18.55 & 7.63 & -- & -- & -- & 1144.16 & 1227.15 & 6.76 \\
& E4 & 54.44 & 61.50 & 11.48 & 18.38 & 18.57 & 1.03 & -- & -- & -- & 861.69 & 958.27 & 10.08 \\
& E5 & 44.15 & 46.98 & 6.03 & 12.42 & 11.43 & 8.71 & -- & -- & -- & 478.12 & 470.06 & 1.72 \\
& E6 & 40.06 & 37.87 & 5.79 & 16.67 & 16.74 & 0.43 & -- & -- & -- & 561.22 & 548.15 & 2.39 \\
\multirow{6}{*}{\textbf{2 L}} \\
& E7 & 56.10 & 56.85 & 1.32 & 24.90 & 24.33 & 2.37 & 50.7 & 51.8 & 2.1 & 1207.92 & 1271.88 & 5.03 \\
& E8 & 30.00 & 33.96 & 11.65 & 24.33 & 24.60 & 1.10 & 31.0 & 27.21 & 13.9 & 636.60 & 723.87 & 12.06 \\
& E9 & 41.71 & 44.23 & 5.71 & 24.45 & 25.00 & 2.21 & 41.8 & 39.9 & 4.8 & 891.85 & 954.37 & 6.55 \\
& E10 & 49.08 & 47.72 & 2.84 & 24.45 & 25.43 & 3.87 & 46.3 & 39.19 & 18.1 & 1071.36 & 1066.28 & 0.48 \\
\bottomrule
\end{tabular}%
}
\end{table}

The results exhibit excellent agreement with the experimentally derived features, with most relative errors ranging from 1–10\% and only a few cases exceeding this range. In practice, such deviations can be considered negligible, or at least acceptable, in the design process, as they are comparable to the intrinsic uncertainties of the printing process and the tolerances on geometric conformity. This confirms the robustness of the approach and is particularly significant given the additional uncertainty introduced by the measurement process when extracting experimental cross-sections. Notably, the predicted contact length is also well captured and could be potentially exploited to gain additional valuable insights into the stability of the printed structure.

These findings highlight both the strength of the Fourier descriptor parameterization and the advantages of training on a synthetically generated dataset that densely and uniformly samples the input parameter space. The model’s ability to reproduce experimental outcomes also validates the PFEM simulations as a reliable source of training data. Overall, \textit{ShapeGen3DCP} delivers accurate and robust predictions, independent of specific printing processes or materials, supporting its safe use in real-world applications.

\section{Discussion on validity ranges and filtering of failure scenarios}\label{sec:discussion}
As discussed in Section \ref{sec:dataset}, \textit{ShapeGen3DCP} can, in principle, provide reliable predictions when the combination of input material and process parameters lies within the dimensional training space. By transforming the input variables and aggregating certain parameters into dimensionless groups, the model’s predictive capacity is further enhanced. This transformation extends the input space, enabling predictions for cases not explicitly represented in the dimensional dataset, provided that, when expressed in terms of these dimensionless parameters, they fall within the learned domain.  

\begin{figure}[h]
			\centering
			\subfloat[Slugs formation]{\includegraphics[width=0.6\linewidth]{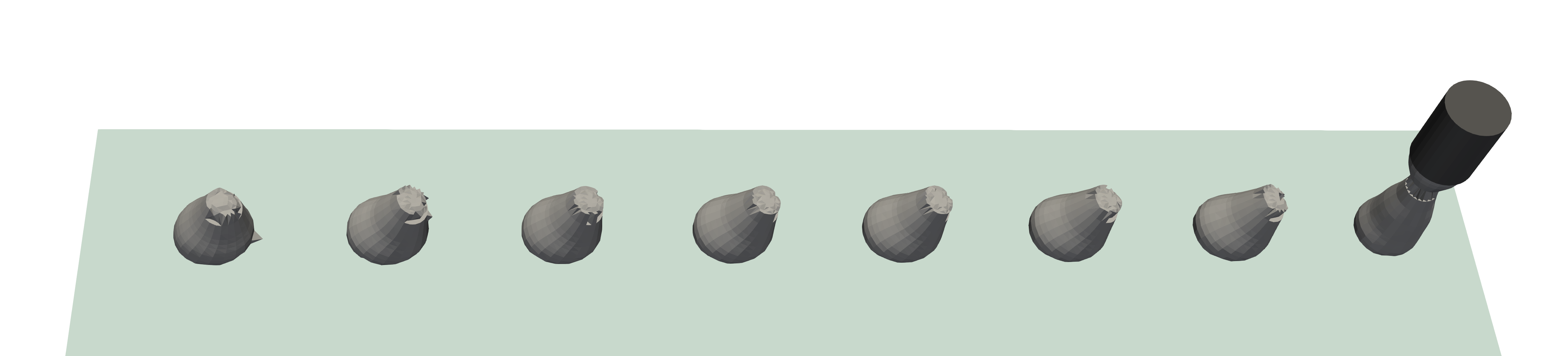}}\\
            \subfloat[Filament tearing]{\includegraphics[width=0.6\linewidth]{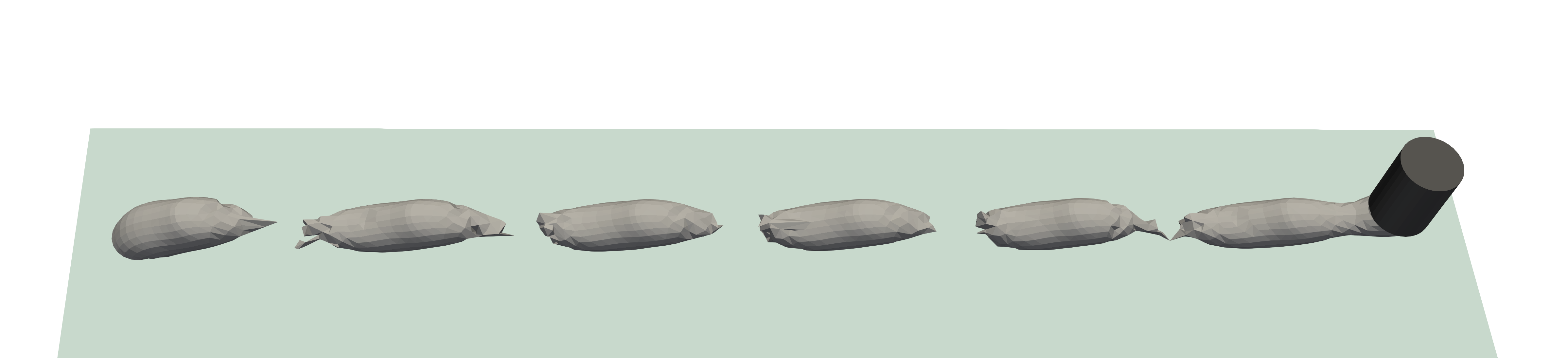}}\\
  		\subfloat[Filament buckling]{\includegraphics[width=0.6\linewidth]{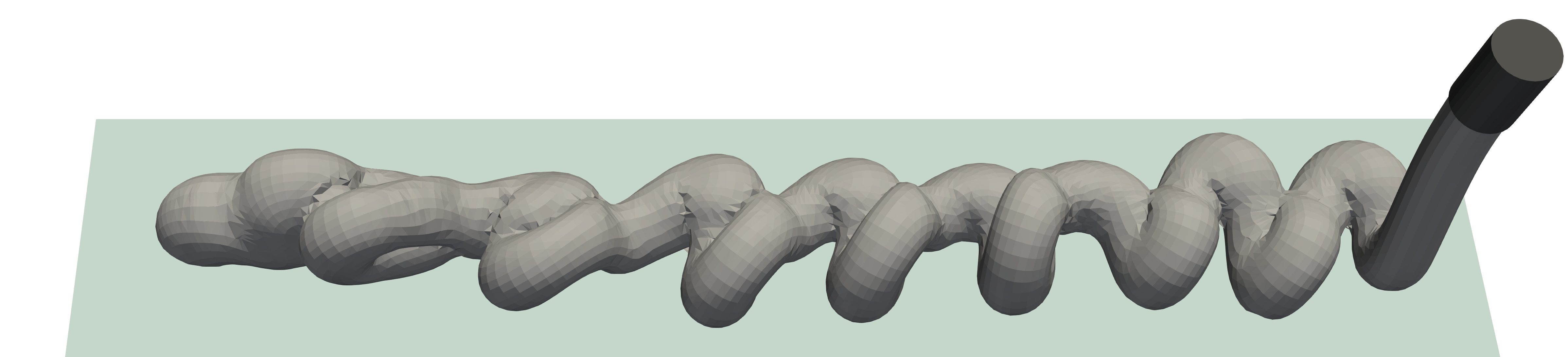}} 
			\caption{Filament failure modes reproduced using the PFEM numerical framework introduced in \cite{rizzieri2023,rizzieri2025a} and adopted in this study.}
			\label{fig:12}
		\end{figure}

The dataset ranges were deliberately chosen to exclude extreme cases, while still being broad enough to encompass a variety of common printing strategies and setups. Nevertheless, certain parameter combinations may still result in unwanted failure modes, such as filament tearing, filament buckling, and slug formation, as illustrated in Figure \ref{fig:12}. Although predicting these scenarios is not the primary goal of this work, the practical aim of developing a ready-to-use tool for practitioners makes it important to issue warnings when such situations are likely to occur.

Below, the integration of empirical failure and stability checks from recent experimental studies is outlined, showing how they could complement the tool’s predictions. However, these empirical checks come with notable limitations and limited accuracy. In the future, it may therefore be convenient to train a dedicated machine learning model to specifically classify and filter out potential failure scenarios.
 
\begin{itemize}

    \item \textbf{Slugs formation}: When the nozzle height exceeds by far the nozzle diameter ($h_n>>\phi_n$), the extruded filament may tear vertically under its own weight, producing discontinuous slugs (Figure \ref{fig:12}-a). In the specific, slug formation will start when a certain critical nozzle height is overcome, i.e., in dimensional terms when: 
    
    \begin{equation}
        h^* > h_c^* = \frac{h_c}{\phi_n},
    \end{equation} 
    
    \noindent where the dimensional critical height $h_c$ can be estimated, according to \cite{geffrault2023}, as: 
    \begin{equation}
        h_c \approx \frac{\sigma_c}{\rho g} \, \frac{r_c}{r_n} 
        - \frac{4}{3}\frac{r_n^2}{r_c} + 6r_n.
    \end{equation}
    \noindent The nozzle radius is defined as $r_n = \phi_n / 2$. The critical radius, corresponding to the filament radius at the transition between the solid and the fluid regimes, can be estimated to be in the range of $r_c \in [0.8 \ r_n , \ 0.9 \ r_n]$ for different yield stress fluids \cite{geffrault2023}. Moreover, the critical stress can be computed as $\sigma_c = \xi \sqrt{3} \tau_0$, where $\xi \simeq 1.5$ is an empirical correction coefficient \cite{geffrault2023a}.

    \item \textbf{Filament tearing}: When the printing velocity significantly exceeds the flow velocity, horizontal tearing of the deposited filament may occur (Figure \ref{fig:12}-b). Although accurately predicting this phenomenon is complex, a few simplified criteria have been proposed. According to \cite{wolfs2021}, tearing is expected when:
    \begin{equation}
        \frac{G}{\tau_0}\ln v^* \gg 1,
    \end{equation}
    where $G$ is the elastic shear modulus of the material and $\tau_0$ the yield stress. Alternatively, \cite{geffrault2023} proposed that tearing occurs when: 
    \begin{equation}
        v^* > (1 - \varepsilon_c)^{-2},
    \end{equation}
    with the critical strain that can be computed as $\varepsilon_c = \frac{\xi\sqrt{3} \tau_0}{G}$, again with $\xi \simeq 1.5$. While useful for providing a first estimate of when tearing may occur, both approaches require knowledge of the elastic shear modulus,which may not be available or easily measured in the fresh state. Furthermore, they neglect the influence of the nozzle height, which can substantially shift the onset of tearing, as hinted in \cite{wei2024a}.
    
    \item \textbf{Filament buckling}:  
    When the nozzle height exceeds the nozzle diameter ($h_n > \phi_n$) and the printing velocity is smaller than the flow velocity ($v_p < u_f$), the extruded filament may buckle and fold onto itself (Figure \ref{fig:12}-c). The onset of this transition, from a straight filament to meanders or loops, can be evaluated using the simplified criterion proposed in \cite{geffrault2023}, which assumes no dependence on the material's rheology and predicts buckling when:
        \begin{equation}
            v^* < 1 - \frac{1}{h^*},
        \end{equation}  
    \noindent where the dimensionless height is defined as $h^* = \frac{h_n}{\phi_n}$.

\end{itemize}

All these checks are integrated into the online version of \textit{ShapeGen3DPC}, which issues an error message whenever the selected input parameters would result in filament tearing, buckling, or slug formation.

\section{Conclusions}\label{sec:conclusions}
This work has introduced \textit{ShapeGen3DCP} (\url{https://www.dica.polimi.it/ai3dcp}), a novel deep learning framework for fast and accurate prediction of cross-sectional geometries of 3D-printed cementitious filaments. Compared to existing approaches, the framework presents several key innovations:

\begin{itemize}
    \item First, instead of predicting isolated geometric features, it learns the full boundary contour of the cross-section. This is achieved through Fourier descriptors, which can provide a compact yet expressive representation of the complex layer's free-surface while ensuring a physically plausible, smooth, and closed geometry.

    \item Second, \textit{ShapeGen3DCP} is the first tool of its kind to incorporate material parameters (density, yield stress, viscosity) in addition to process parameters (nozzle diameter, nozzle height, print velocity, flow velocity). This is made possible by leveraging a synthetically generated dataset based on the advanced Particle Finite Element Method (PFEM) model of 3DCP \cite{rizzieri2023,rizzieri2025}, which enables systematic exploration of a wide range of material-process combinations and thus an extensive and uniform coverage of the input space.

    \item By grouping selected input quantities into dimensionless parameters based on physical principles, the model extends its generalization capability beyond the dimensional training set, enabling accurate predictions even for unseen parameter combinations, provided that the corresponding dimensionless values are represented in the training space.
\end{itemize}

Overall, the results confirm the exceptional predictive accuracy and robustness of the proposed approach, which has been validated with experimental data from multiple research groups. Compared to conventional simulation methods (e.g., FEM, PFEM, meshfree methods), \textit{ShapeGen3DCP} delivers cross-sectional predictions with orders-of-magnitude gains in computational efficiency, producing results in milliseconds rather than hours. Combined with its user-friendly implementation as an open web application, the framework demonstrates strong potential for practical adoption.

For designers, early knowledge of filament geometry enables the achievement of improved dimensional conformity, reduced reliance on trial-and-error calibration, and optimized toolpath strategies. The ability to predict interlayer contact also offers valuable insights into adhesion quality and structural integrity, enabling integration with layer-activation FEM 3DCP simulations to enhance geometric precision.

There remains, however, a significant scope for improvement. At present, filament failure phenomena such as tearing or buckling are captured only through simplified empirical checks, and the model outputs purely geometrical information. Future work should extend predictions to additional quantities of interest (e.g., stresses, pressures) and generalize the framework beyond circular nozzles and rectilinear toolpaths. As experimental databases for 3DCP continue to expand \cite{versteege2025,robens-radermacher2025}, increasingly accurate data will become available to partly or fully replace the synthetic dataset used in this study. Owing to its generality, the proposed framework can be readily adapted to train new models on such experimentally enriched databases.

In conclusion, \textit{ShapeGen3DCP} marks a significant step toward computationally efficient and physically informed prediction of 3D concrete printing outcomes. With further development, such as incorporating real-time sensor data for closed-loop control, it could evolve into a key enabler of digital twins for additive manufacturing in construction, bridging high-fidelity simulation and real-world application to enhance both efficiency and reliability in digital fabrication.

\section*{Acknowledgements}
The first and third authors acknowledge the support of the MUSA - Multilayered Urban Sustainability Action - project, funded by the European Union - NextGenerationEU, under the National Recovery and Resilience Plan (NRRP) Mission 4 Component 2 Investment Line 1.5: Strengthening of research structures and creation of R\&D “innovation ecosystems”, set up of “territorial leaders in R\&D''. This research was also partially supported by the Italian Ministry of University and Research, through the project PRIN2022 DTWIX: development of Digital TWIns for multiphysics simulation of eXtreme events in civil engineering (PRIN DTWIX - 2022AL5MSN), and by ICSC—Centro Nazionale di Ricerca in High Performance Computing, Big Data, and Quantum Computing funded by European Union—NextGenerationEU.

\bibliographystyle{elsarticle-num.bst}
\bibliography{Bibliography}  






\end{document}